%% file: brainstorm_paper.tex
\documentclass[%
 reprint,
superscriptaddress,
showkeys,
nofootinbib,
 amsmath,amssymb,
 aps,
prd,
]{revtex4-2}

\input{preamble.tex}
\usepackage{graphicx}% Include figure files
\usepackage{dcolumn}% Align table columns on decimal point
\usepackage{bm}% bold math
\usepackage{hyperref}% add hypertext capabilities
\usepackage{orcidlink}
\usepackage{xstring}

\usepackage{float}
\usepackage{caption}
\usepackage{subcaption}
\usepackage{amsmath,empheq,eqparbox,cases}
\DeclarePairedDelimiter\abs{\lvert}{\rvert}%
\DeclarePairedDelimiter\norm{\lVert}{\rVert}%
\makeatletter
\let\oldabs\abs
\def\abs{\@ifstar{\oldabs}{\oldabs*}}
\let\oldnorm\norm
\def\norm{\@ifstar{\oldnorm}{\oldnorm*}}
\makeatother
\usepackage{braket}

\newcommand{\qb}{\textit{q}\textsc{Bounce}\xspace}

\newcommand{\bc}[3][0]{\left. \phantom{\frac{1}{2}} #3 \right\lvert_{#2=#1}}

\newcommand{\deriv}[3]{%
  \IfEqCase{#1}{%
    {1}{\frac{d#3}{d#2}}% Case for first derivative
  }[\frac{d^{#1}#3}{d#2^{#1}}]% Default case for nth derivative
}

\def \ee{\end{equation}}
\def \be{\begin{equation}}
\def \eea{\end{eqnarray}}
\def \bea{\begin{eqnarray}}

\begin{document}

\title{Generalized Boundary Conditions for the \qb Experiment}

% \title{Generalized Boundary Conditions for Ultracold Neutron Experiments}

% \title{Respecting Boundaries: Boundary Effects on the Linear Gravitational Potential and Implications to \qb}

% \title{Generalized Boundary Conditions for the \qb Experiment and Towards Resolving Experimental Conundrum}

% \title{Quantum Bouncing Ball with Generalized Boundary Conditions: Implications for \qb Experiments}

\author{Eric J. Sung\orcidlink{0000-0003-4437-1068}}
\email{jsung1@arizona.edu}
% \affiliation{University of Arizona, Tucson, AZ 85721, USA}
\affiliation{Program in Applied Mathematics, University of Arizona, Tucson, AZ 85721, USA}
% \affiliation{Graduate Interdisciplinary Program in Applied Mathematics, University of Arizona, Tucson, AZ 85721, USA}
\affiliation{Tulane University, New Orleans, LA 70118, USA}
% \author{Eric J. Sung\orcidlink{0000-0003-4437-1068}}
% \affiliation{Tulane University, New Orleans, LA 70118, USA}
% \email{jsung2@tulane.edu}

\author{Benjamin Koch\orcidlink{0000-0002-2616-0200}}
\email{benjamin.koch@tuwien.ac.at}
\affiliation{Institut f{\"u}r Theoretische Physik, Technische Universit{\"a}t Wien, Wiedner Hauptstrasse 8–10, A-1040 Vienna, Austria}
\affiliation{Pontificia Universidad Cat{\'o}lica de Chile
Instituto de F{\'i}sica, Pontificia Universidad Cat{\'o}lica de Chile,
Casilla 306, Santiago, Chile}

\author{Tobias Jenke\orcidlink{0000-0002-7815-4726}}
\email{jenke@ill.fr}
\affiliation{Institut Laue-Langevin, 71 avenue des Martyrs, 38000 Grenoble, France}

\author{Hartmut Abele\orcidlink{0000-0002-6832-9051}}
\email{hartmut.abele@tuwien.ac.at}
\affiliation{Technische Universitat Wien, Atominstitut, Stadionallee 2, 1020 Wien, Austria}

\author{Denys I. Bondar\orcidlink{0000-0002-3626-4804}}
\email{dbondar@tulane.edu}
\affiliation{Tulane University, New Orleans, LA 70118, USA}

%\affiliation{%


%\affiliation{
%\affiliation{
%\affiliation{%


\date{\today}% It is always \today, today,

\begin{abstract}
Discrepancies between theory and recent \qb\ data have prompted renewed scrutiny of how boundary conditions are implemented for ultracold neutrons bouncing above a mirror in Earth’s gravity. We apply the theory of self-adjoint extensions to the linear gravitational potential on the half-line and derive the most general boundary condition that renders the Hamiltonian self-adjoint. This introduces a single real self-adjoint parameter \(\lambda\) that continuously interpolates between the Dirichlet case and more general (Robin-type) reflecting surfaces. 

Building on this framework, we provide analytical expressions for the energy spectrum, eigenfunctions, relevant matrix elements, and a set of sum rules valid for arbitrary \(\lambda\). We show how nontrivial boundary conditions can bias  measurements of \(g\) and can mimic or mask putative short-range ``fifth-force''. %thereby offering a boundary-based explanation for the reported tension. We further outline a strategy to enhance the sensitivity of \qb\ by exploiting boundary engineering, turning \(\lambda\) into a tunable control knob.
Our results emphasize that enforcing self-adjointness—and modeling the correct boundary physics—is essential for quantitative predictions in gravitational quantum states. %We argue that a careful treatment of \(\lambda\) is necessary before attributing anomalies to new interactions, and we point to upcoming numerical fits to recent \qb\ measurements aimed at extracting \(\lambda\) directly from data. 
Beyond neutron quantum bounces, the approach is broadly applicable to systems where boundaries and self-adjointness govern the observable spectra and dynamics.
%\begin{description}
%\end{description}
\end{abstract}

\maketitle

%\tableofcontents

\renewcommand\theequation{\arabic{section}.\arabic{equation}}
\counterwithin*{equation}{section}

\section{Introduction} \label{section: intro}

% \textcolor{blue}{(07/31/2025) Note to readers: Any text highlighted in blue is likely to be commented out. I have only kept it because it seems like it could be potentially important.}

% \textcolor{blue}{Note to Denys (2/27/23): I noticed in Hartmut's paper \cite{micko_qbounce_2023} that he used $g_{c}=9.804 925 \, \text{m}/\text{s}^{2} \approx 9.805 \, \text{m}/\text{s}^{2}$ for the local gravitational acceleration. Is this the exact value of the local gravitational acceleration at the Institut Laue-Langevin? For our work, I used the SI standard $g_{c}=9.80665 \, \text{m}/\text{s}^{2}$. If I use Hartmut's value of $g_{c}$, I get $\lambda =  0.11654029361897301$.} \\

% \textcolor{blue}{Note to Denys (3/3/23): In accordance with our last meeting with Hartmut on 3/1/23, I have switched and fixed all calculations to use $g_{c}=9.804 925 \, \text{m}/\text{s}^{2} \approx 9.805 \, \text{m}/\text{s}^{2}$ for the local gravitational acceleration.} \\

% Boundary conditions in physics are indispensable in understanding phenomena.

In quantum physics, the notion of Hermitian and self-adjoint operators is not usually distinguished and it is often the case that both terms are used interchangeably. This cavalier approach is often perfectly acceptable in many physical problems of interest, but when considered in the context of domains with boundaries, the distinction between Hermitian and self-adjoint becomes crucial. For instance, there arises an interesting issue regarding the self-adjointness of the Hamiltonian operator since the Hamiltonian operator can be Hermitian but not self-adjoint depending on the domain of interest. However, this would make the Hamiltonian operator not a proper observable. Fortunately, there exists a mathematical theory of self-adjoint extensions \cite{von2018mathematical,bonneau_self-adjoint_2001,juricObservablesQuantumMechanics2022} where an operator, under certain conditions, can be systematically extended to a self-adjoint operator by classifying the boundary conditions that ensures self-adjointness on the specified domain. This restores the Hamiltonian as a proper observable at the price of a new general boundary condition. 

This beautiful mathematical theory provides an intimate connection between self-adjoint operators and boundary conditions in addition to revealing new results. Self-adjoint extensions have been applied to a surprisingly wide variety of fields such as confined particles \cite{al-hashimi_particle_2012,al-hashimi_self-adjoint_2012}, Aharonov-Bohm effect \cite{gerbertFermionsAharonovBohmField1989,gavrilovDiracEquationMagneticsolenoid2004,salemSelfAdjointExtensionApproach2019}, heavy atoms \cite{voronovDiracHamiltonianSuperstrong2007, voronovPeculiaritiesElectronEnergy2016}, black holes \cite{govindarajanHorizonStatesAdS2000,birminghamNearhorizonConformalStructure2001, guptaFurtherEvidenceConformal2002, guptaNearHorizonConformalStructure2002,balachandranNearHorizonModesSelfadjoint2019,gielenQuantumSchwarzschildAdSBlack2025}, anomalies \cite{esteveAnomaliesConservationLaws1986,esteveOriginAnomaliesModified2002,guptaAnomaliesRenormalizationMixed2014}, spontaneously broken symmetries \cite{capriSelfadjointnessSpontaneouslyBroken1977,bermanBoundaryEffectsQuantum1991}, Weyl semimetals \cite{seradjeh_surface_2018}, topological phases \cite{tanhayiahariRoleSelfadjointnessContinuum2016a}, and entropy conservation in classical systems \cite{mccaul_entropy_2019}. More recently in Ref.~\cite{albrecht_bouncing_2023}, self-adjoint extensions were used for the particle in a box Hamiltonian by using a new self-adjoint momentum operator introduced in Refs.~\cite{al-hashimi_canonical_2021,al-hashimi_alternative_2021}, leading to a new energy spectrum. The novel results that self-adjoint extensions have yielded naturally raises the question of whether or not this theory can be experimentally verified. We believe that self-adjoint extensions may provide physical results and could potentially affect the interpretation of past experimental data as well.

In this paper, we apply the theory of self-adjoint extensions to the quantum linear potential to model the recent \cite{micko_qbounce_2023,mickoQBounceRamseySpectroscopy2023} quantum bouncing ball, or \qb \cite{nesvizhevsky_quantum_2002,nesvizhevsky_measurement_2003,cronenberg_acoustic_2018,micko_qbounce_2023,mickoQBounceRamseySpectroscopy2023}, experiment for ultracold neutrons (UCNs) performed at the instrument
PF2 at the Institut Laue-Langevin (ILL) in Grenoble, France. In the latest \qb experiment \cite{micko_qbounce_2023,mickoQBounceRamseySpectroscopy2023}, the $\ket{1} \to \ket{6}$ transition frequency of UCNs was measured via Ramsey Gravity Resonance Spectroscopy \cite{abele_ramseys_2010,rechberger_ramsey_2018, sedmik_proof_2019}. From these measurements, the neutron's local acceleration value $g$ was determined and found to have a slight discrepancy from the 2021 measured local acceleration $g_{c}=9.804 925 \, \text{m}/\text{s}^{2}$. In our model, boundary effects of the mirror led to different local acceleration values. The linear potential is a simple textbook case of an easily solvable Hamiltonian, yet despite its mathematical simplicity, has found applications in a variety of fields other than free-falling UCNs such as the Stark effect \cite{robinettStarkEffectLinear2009}, bouncing Bose–Einstein condensates \cite{bongsCoherentEvolutionBouncing1999}, and the energy spectrum of quarkonium \cite{bhanotNewPotentialQuarkonium1978}. This diverse range of applications naturally makes the linear potential an excellent contender for self-adjoint extensions since new insights in one field of application can lead to insights in the other aforementioned fields.

% the energy spectrum of quarkonium, quantum wells in semiconductor heterostructures, and Bloch oscillations in crystals. This diverse range of applications naturally makes the linear potential an excellent contender for self-adjoint extensions since new insights in one field of application can be lead to insights in the aforementioned fields. 

In Sec.~\ref{sec:quantum_bouncing_ball}, we first provide a brief overview of the linear potential and solve it for the Dirichlet and Neumann boundary conditions. Then we present a brief overview of self-adjoint extensions and general boundary conditions in Sec.~\ref{sec:self_adjoint_operators} and apply it to the linear potential to find the most general boundary conditions in Sec.~\ref{sec:general_boundary_conditions}. Next, we solve the resulting eigenvalue problem in Sec.~\ref{sec:generalized_eigenenergies_and_eigenfunctions} and derive the generalized Ehrenfest theorems in Sec.~\ref{sec:generalized_ehrenfest_theorems}. Then we use our previous results to derive new closed expressions of the generalized matrix elements and sum rules in Secs.~\ref{sec:exact_matrix_elements} and \ref{sec:generalized_sum_rules}, respectively. In Sec.~\ref{sec:generalized_heisenberg_uncertainty_rel}, we demonstrate that the Heisenberg uncertainty relation changes due to the general boundary condition. In Sec.~\ref{sec:applications}, we apply our results to the recent \cite{micko_qbounce_2023,mickoQBounceRamseySpectroscopy2023} \qb experiment and demonstrate that the general boundary condition can lead to nontrivial boundary effects that influence the measurement of the local gravitational acceleration. Then we close our discussion with an outlook on the applications and impact of our work in Sec.~\ref{sec:discussion and outlook}.

In this paper, we use $g_{c}=9.804 925 \, \text{m}/\text{s}^{2}$ to denote the 2021 local gravitational acceleration measured at the ILL. We reserve $g$ to denote the experimentally measured local gravitational acceleration. The binary operations $[\cdot,\cdot]$ and $\{\cdot,\cdot \}$ denote the commutator and anticommutator, respectively.

\section{Quantum Linear Potential}\label{sec:quantum_bouncing_ball}

We begin with a brief review of the quantum linear potential for the Dirichlet and Neumann boundary conditions. Although this problem is solved in many textbooks (see, e.g.,~\cite{sakurai2020modern}), it proves useful in having the results at our disposal as well as showing the uncommon case of the Neumann condition.

Recall that the linear potential Hamiltonian is 
\begin{align}
    \hat{H} &= \frac{\hat{p}^{2}}{2m} + F_{0}\abs{\hat{x}}, \label{linear_pot_ham}
\end{align}
where $F_{0} \in [0,\infty)$ is a constant with units of force. Hamiltonian \eqref{linear_pot_ham} admits the eigenvalue equation $\hat{H}\ket{E}=E\ket{E}$ which in the coordinate representation is
% The Hamiltonian \eqref{linear_pot_ham} has the eigenvalue equation 
% \begin{align}
%     \hat{H} \ket{E} &= E \ket{E}, \label{qbounce_eig_eq}
% \end{align}
% which in the coordinate representation is
\begin{align}
    \frac{d^{2}\psi}{dz^{2}} = z \psi, \label{qbounce_diff_eq}
\end{align}
where we have defined the dimensionless variables
\begin{align}
    z = \xi - \varepsilon,  \quad \xi = \frac{x}{x_{0}}, \quad \varepsilon = \frac{E}{\mathcal{E}_{0}},
\end{align}
along with the characteristic length and energy
\begin{gather}
    x_{0} = \left( \frac{\hbar^{2}}{2mF_{0}} \right)^{1/3}, \quad \mathcal{E}_{0} = F_{0}x_{0} = \left( \frac{\hbar^{2} F_{0}^{2}}{2m} \right)^{1/3}, \label{non_dim_var}
\end{gather}
respectively. 
% We note that numerically, the characteristic units have the values
% \begin{align}
%     x_{0} = 5.87 \, \mu \text{m}, \quad \mathcal{E}_{0} = 0.6016 \, \text{peV}.
% \end{align}
The only normalizable solution of Eq.~\eqref{qbounce_diff_eq} is the Airy function
\begin{align}
    \braket{x | E} &\equiv \psi(z) = \mathcal{N} \text{Ai} (z), \label{pre_eigenfunc}
\end{align}
where $\mathcal{N}$ is the normalization constant. Note that we have not yet applied a specific boundary condition to the eigenfunction \eqref{pre_eigenfunc}.

If we use the Dirichlet boundary condition $\psi(x=0)=0$, we find that the quantized energy levels are 
\begin{align}
    E^{D}_{n} &= -\mathcal{E}_{0} a_{n}, \quad n=1,2,\ldots, \label{dirichlet_energy}
\end{align}
where $a_{n}$ are the (Dirichlet) roots of $\text{Ai}(x)$. Then we have a complete set of orthonormal eigenfunctions $\{\ket{n^{D}} \}^{\infty}_{n=1}$ 
\begin{align}
    \braket{x | n^{D}} &= \psi^{D}_{n}(\xi) = \mathcal{N}^{D}_{n} \text{Ai} \left( \xi + a_{n} \right), \label{qbounce_eigenfunc_D} \\
    \mathcal{N}^{D}_{n} &= \frac{1}{\sqrt{x_{0}} \abs{\text{Ai}^{\prime}(a_{n})}}. \label{diri norm}
\end{align}
If we use the Neumann boundary condition $\psi^{\prime}(x=0)=0$, we find that the quantized energy levels are 
\begin{align}
    E^{N}_{n} &= -\mathcal{E}_{0} a^{\prime}_{n}, \quad n=1,2,\ldots, \label{neumann_energy}
\end{align}
where $a^{\prime}_{n}$ are the (Neumann) roots of $\text{Ai}^{\prime}(x)$. Then we have a complete set of eigenfunctions $\{\ket{n^{N}} \}^{\infty}_{n=1}$
\begin{align}
    \braket{x | n^{N}} &= \psi^{N}_{n}(\xi) = \mathcal{N}^{N}_{n} \text{Ai} \left( \xi + a^{\prime}_{n} \right), \label{qbounce_eigenfunc_N} \\
    \mathcal{N}^{N}_{n} &= \frac{1}{\sqrt{x_{0} \abs{a^{\prime}_{n}}} \abs{\text{Ai}(a^{\prime}_{n})}}. \label{neu norm}
\end{align}
Note that we have adopted the negative sign convention for $a_{n}$ and $a_{n}'$ $(a_{n}, a_{n}' < 0)$ which we will adopt throughout the paper. 

\section{The Theory of Self-Adjoint Operators and Extensions} \label{sec:self_adjoint_operators}

In this section, we provide a brief review of self-adjoint extensions based on Refs.~\cite{bonneau_self-adjoint_2001,araujo_operator_2004,mccaul_entropy_2019,juricObservablesQuantumMechanics2022}. We will utilize this knowledge to construct the general boundary condition for the linear potential \eqref{linear_pot_ham}.

Let $\mathcal{H}$ be a Hilbert space. An operator defined on $\mathcal{H}$ is a pair $(\hat{A}, \mathcal{D}(\hat{A}))$ where $\hat{A}$ and $\mathcal{D}(\hat{A})$ are the linear mapping $\hat{A}: \mathcal{D}(\hat{A}) \to \mathcal{H}$ and the domain of $\hat{A}$, respectively. An operator $\hat{A}$ is \textit{densely defined} if $\mathcal{D}(\hat{A})$ is dense in $\mathcal{H}$. For a densely defined operator $\hat{A}$, its adjoint $\hat{A}^{\dagger}$ is defined as follows: a vector $\phi \in \mathcal{H}$ belongs to $\mathcal{D}(\hat{A}^{\dagger})$ if and only if there exists a unique $z\in\mathcal{H}$ such that
\begin{align}
    \braket{\phi|\hat{A}\psi} = \braket{z|\psi}, \quad \forall \psi \in \mathcal{D}(\hat{A}),
\end{align}
or equivalently, the linear functional $f_{\phi}(\psi) \coloneq \braket{\phi| \hat{A} \psi}$ is continuous. Then one has
\begin{align}
    \braket{\phi|\hat{A}\psi} = \braket{\hat{A}^{\dagger}\phi|\psi}, \quad \forall \phi \in \mathcal{D}(\hat{A}^{\dagger}), \, \forall \psi \in \mathcal{D}(\hat{A}).
\end{align}
It immediately follows that for any densely defined operator, its adjoint is \textit{closed} \cite{akhiezer2013theory}. An operator $\hat{A}$ is \textit{Hermitian} (or \textit{symmetric}) if $\mathcal{D}(\hat{A}) \subseteq \mathcal{D}(\hat{A}^{\dagger})$ and
\begin{align}
    \braket{\phi|\hat{A}\psi} = \braket{\hat{A}\phi|\psi}, \quad \forall \phi,\psi \in \mathcal{D}(\hat{A}).
\end{align}
An operator $\hat{A}$ is \textit{self-adjoint} if $\mathcal{D}(\hat{A})=\mathcal{D}(\hat{A}^{\dagger})$. Henceforth, we will assume that $(\hat{A}, \mathcal{D}(\hat{A}))$ is densely defined, Hermitian, and closed with $(\hat{A}^{\dagger}, \mathcal{D}(\hat{A}^{\dagger}))$ as its adjoint. 

It is not always the case that the Hermitian operator $\hat{A}$ is self-adjoint as well. Consider the general Hamiltonian 
\begin{align}
    \hat{H}_{0} = \frac{\hat{p}^{2}}{2m} + U(\hat{x}).
\end{align}
The operator $\hat{H}_{0}$ is Hermitian but not necessarily self-adjoint. To see this, symmetry on $\mathcal{D}(\hat{H}_{0})$ yields
\begin{align}
    &\braket{\hat{H}_{0}\varphi_{2} | \varphi_{1}} - \braket{\varphi_{2} | \hat{H}_{0}\varphi_{1}} \notag \\
    &= \frac{\hbar^{2}}{2m} \left.\left[ \varphi_{1} \frac{d \varphi^{\ast}_{2}}{dx} - \varphi^{\ast}_{2} \frac{d \varphi_{1}}{dx} \right] \right\lvert_{\text{boundary}},\label{self_adjoint_cond}
\end{align}
for any $\phi_{1},\phi_{2} \in \mathcal{D}(\hat{H}_{0})$ where on the half-line $[0,\infty)$, the boundary reduces to $x=0$. Thus, self-adjointness hinges on choosing $\mathcal{D}(\hat{H}_{0})$ such that the boundary term in Eq.~\eqref{self_adjoint_cond} vanishes for all $\phi_{1},\phi_{2} \in \mathcal{D}(\hat{H}_{0})$. Fortunately, the \textit{von Neumann Deficiency Index Theorem} \cite{von2018mathematical} yields the criteria for whether operator $\hat{A}$ can be made self-adjoint. More specifically, the deficiency index theorem determines all possibilities for $\hat{A}$ with domain $\mathcal{D}(\hat{A})$ by considering the eigenstates $\psi_{\pm}$ in $\mathcal{D}(\hat{A}^{\dagger})$:
\begin{align}
    \hat{A}^{\dagger} \psi_{\pm} &= \pm i\eta \psi_{\pm}, \quad \eta \in \mathbb{R}. \label{self adjoint eig eq}
\end{align}
The number of linearly independent solutions for $\psi_{\pm}$ are the \textit{deficiency indices} $n_{\pm}$. For any operator $\hat{A}$ with deficiency indices $(n_{+},n_{-})$, we have the three possibilities:
\begin{enumerate}
    \item $n_{+}=n_{-}=0$: $\hat{A}$ is self-adjoint,

    \item $n_{+}=n_{-}=n\geq 1$: $\hat{A}$ has a $U(n)$-family of self-adjoint extensions where $U(n)$ is the group of $n\times n$ unitary matrices,

    \item $n_{+}\neq n_{-}$: $\hat{A}$ has no self-adjoint extension. 
\end{enumerate}
If $\hat{A}$ has a self-adjoint extension, we can find the boundary condition by constructing the general solution
\begin{align}
    \tilde{\psi}(x) = \psi_{+}(x) + \alpha \psi_{-}(x), \quad \alpha \in U(n),
\end{align}
and evaluating Eq.~\eqref{self_adjoint_cond} with $\varphi_{1}(x)=\psi(x)$ and $\varphi_{2}(x)=\tilde{\psi}(x)$.

To summarize this section, when a Hermitian operator is not self-adjoint on some specified domain of interest, the theory of self-adjoint extensions systematically classifies all boundary conditions that make the operator self-adjoint \textit{on that particular domain}. This in turn assures a real spectrum, unitary time evolution, and probability current conservation on the particular domain. 

\section{General Boundary Condition}\label{sec:general_boundary_conditions}

In this section, we derive the general boundary condition for the linear potential \eqref{linear_pot_ham} based on the theory of self-adjoint operators developed in Sec.~\ref{sec:self_adjoint_operators}. Since the wave function $\psi(x)$ must vanish at $+\infty$, we only consider the boundary condition at $x=0$. 

As shown in Sec.~\ref{sec:self_adjoint_operators}, we seek to find the solutions to the equation
\begin{align}
    \hat{H} \psi_{\pm} &= \pm i\eta \psi_{\pm}, \quad \eta \in \mathbb{R}. \label{self_adj_ham}
\end{align}
We first use the non-dimensional variables
\begin{align}
    z_{\pm} = \xi \mp i\varepsilon_{\eta}, \quad \varepsilon_{\eta} = \frac{\eta}{\mathcal{E}_{0}},
\end{align}
in Eq.~\eqref{self_adj_ham} to get
\begin{align}
    \frac{d^{2}\psi_{\pm}}{dz^{2}_{\pm}} &= z_{\pm} \psi_{\pm}. \label{selfadj_diffeq}
\end{align}
In the linear potential's Hilbert space $\mathcal{H}=L^{2}((0,+\infty))$, the only normalizable solutions of Eq.~\eqref{selfadj_diffeq} are
\begin{align}
    \psi_{\pm}(\xi) &=  \text{Ai}(\xi \mp i\varepsilon_{\eta}).
\end{align}
Thus the deficiency indices are $(n_{+}=1,n_{-}=1)$ so by the deficiency index theorem, Hamiltonian \eqref{linear_pot_ham} has a family of self-adjoint extensions parameterized by a single complex phase, i.e., $U(1)$.
% \begin{align}
%     U(1) &= \left\{ \alpha \in \mathbb{C} \, \big\lvert \, \abs{\alpha}=1 \right\}.
% \end{align}

To find the general boundary condition, we choose an arbitrary $\alpha \in U(1)$ which has the form
\begin{align}
    \alpha = e^{i\theta}, \quad \theta \in [0,2\pi],
\end{align}
and define 
\begin{align}
    \tilde{\psi}(x) &= \psi_{+}(x) + \alpha \psi_{-}(x).
\end{align}
Then we use Eq.~\eqref{self_adjoint_cond} with Hamiltonian \eqref{linear_pot_ham}, $\varphi_{1}(x)=\psi(x)$, and $\varphi_{2}(x)=\tilde{\psi}(x)$ to get
\begin{align}
    &\psi(0) \left[ \text{Ai}^{\prime}(i\varepsilon_{\eta}) + \alpha^{\ast} \text{Ai}^{\prime}(-i\varepsilon_{\eta}) \right] \notag \\
    &- \left[ \text{Ai}(i\varepsilon_{\eta}) + \alpha^{\ast} \text{Ai}(-i\varepsilon_{\eta}) \right] \frac{d\psi(0)}{d\xi} = 0, \label{pre bc 0}
\end{align}
where the prime denotes differentiation with respect to $\xi$. We can rewrite Eq.~\eqref{pre bc 0} as
\begin{align}
    \frac{1}{\psi(0)}  \frac{d\psi(0)}{d\xi} &= \frac{\text{Ai}^{\prime}(i\varepsilon_{\eta}) + \alpha^{\ast} \text{Ai}^{\prime}(-i\varepsilon_{\eta})}{\text{Ai}(i\varepsilon_{\eta}) + \alpha^{\ast} \text{Ai}(-i\varepsilon_{\eta})} = \lambda, \label{pre_bc}
\end{align}
where $\lambda \in \mathbb{R}$ is the dimensionless self-adjoint parameter (see Appendix~\ref{app:proof_of_lambda_in_real} for a proof of $\lambda \in \mathbb{R}$). Then the general boundary condition for Eq.~\eqref{linear_pot_ham} is
\begin{align}
    \psi(\xi) \big\lvert_{\xi=0} &= \lambda \psi^{\prime}(\xi) \big\lvert_{\xi=0}, \label{general_bc} 
\end{align}
or, equivalently, with the correct dimensions
\begin{align}
    \left. \vphantom{\frac{1}{1}} \psi(x) \right\lvert_{x=0} &= \left. \lambda x_{0} \frac{d\psi}{dx} \right\lvert_{x=0}, \label{general_bc_units}
\end{align}
which is of the Robin (mixed) type. For each value of $\lambda$, we now get a different self-adjoint operator $\hat{H}_{\lambda}$ that yield different physical results depending on the choice of $\lambda$. Applying Eq.~\eqref{general_bc} to the probability current then yields
\begin{align}
    j(x) \big\lvert_{x=0} = \left. \left[\psi^{\ast} \frac{d\psi}{dx} - \frac{d\psi^{\ast}}{dx} \psi \right] \right\lvert_{x=0} = 0. \label{prob current at bd}
\end{align}
We see that the current \eqref{prob current at bd} will not leak through the boundary $x=0$, thus preserving unitarity on the half line.

\section{Generalized Eigenenergies and Eigenfunctions} \label{sec:generalized_eigenenergies_and_eigenfunctions}

In this section, we solve the eigenvalue problem of Hamiltonian \eqref{linear_pot_ham} using the general boundary condition \eqref{general_bc}. We also provide an approximate formula for the generalized eigenenergies.

\begin{figure}
    % \centering
    \begin{subfigure}[t]{.49\textwidth}
        \includegraphics[width=\columnwidth]{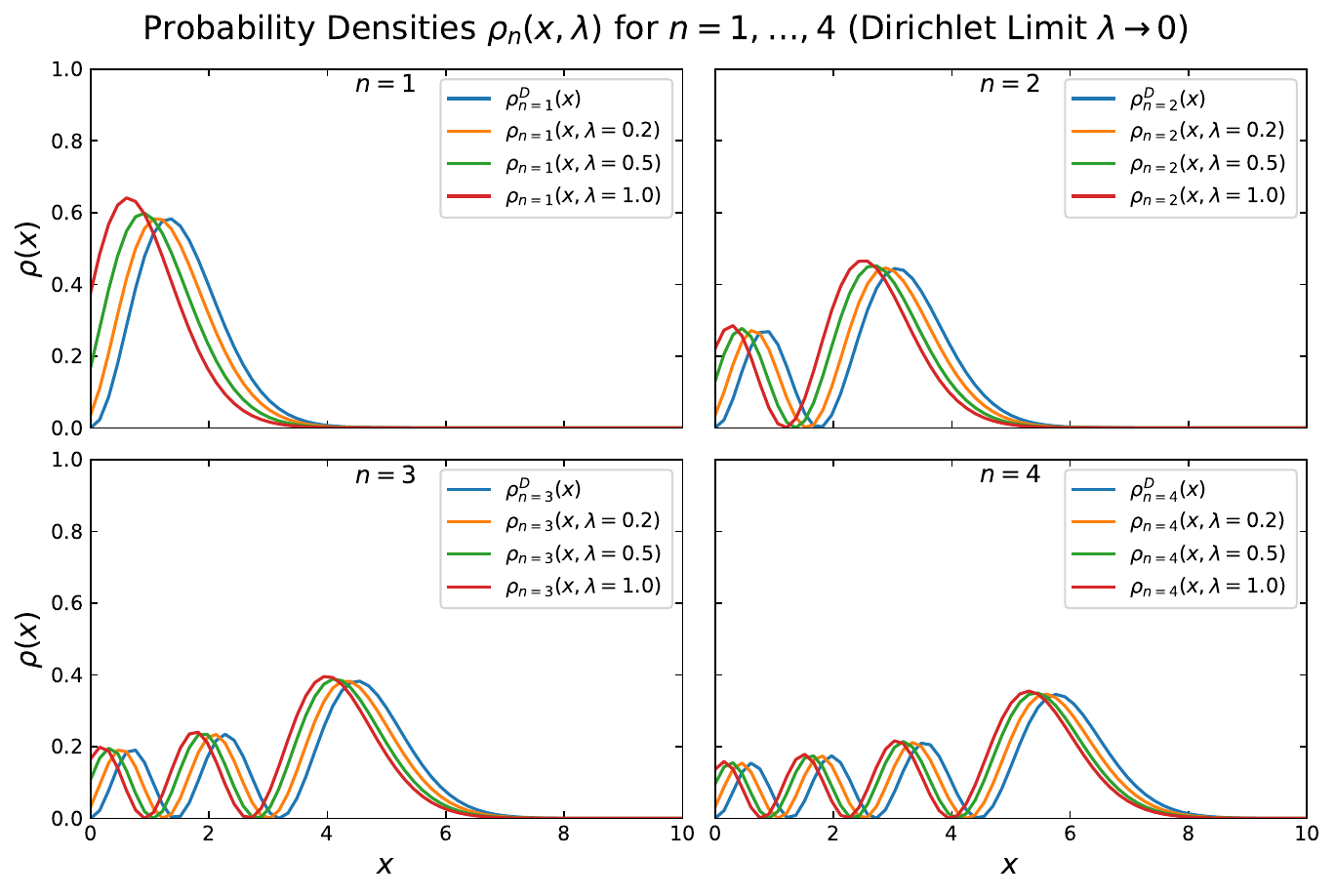}% Here is how to import EPS art
        \caption{\label{fig:psigridplt} General probability density $\rho_{n}(x,\lambda)$ in the limit $\lambda \rightarrow 0$. Note how $\rho_{n}(x,\lambda)$ starts to fit $\rho_{n}^{D}(x)$ with the Dirichlet condition as $\lambda \rightarrow 0$.}
    \end{subfigure} \\
    \begin{subfigure}[t]{.49\textwidth}
        \includegraphics[width=\columnwidth]{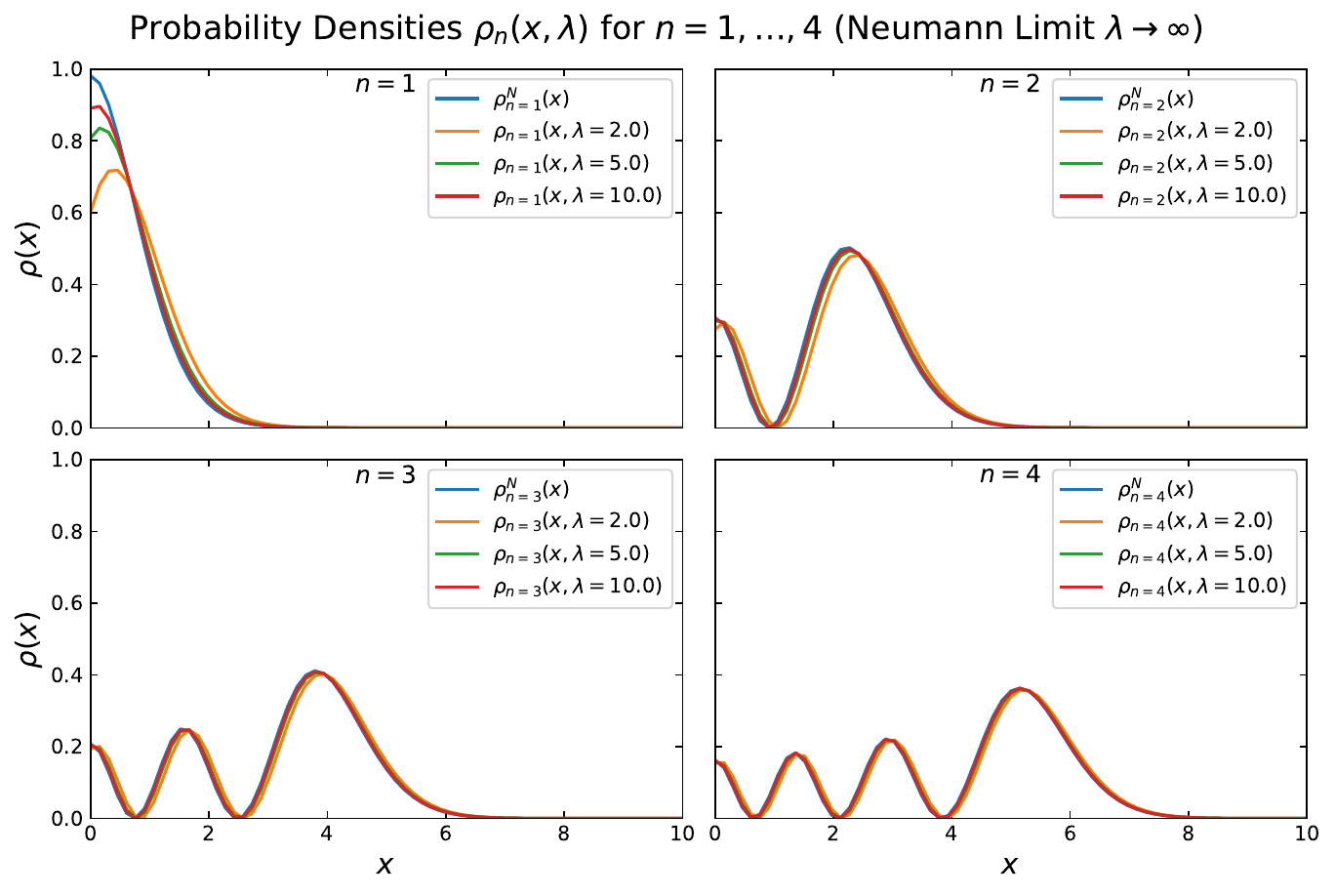}% Here is how to import EPS art
        \caption{\label{fig:psipgridplt} General probability density $\rho_{n}(x,\lambda)$ in the limit $\lambda \rightarrow \infty$. Note how $\rho_{n}(x,\lambda)$ starts to fit $\rho^{N}_{n}(x)$ with the Neumann condition as $\lambda \rightarrow \infty$.}
    \end{subfigure}%
    \caption{Dirichlet $\rho_{n}^{D}(x) = \lvert \psi_{n}^{D}(x) \rvert^{2}$ and Neumann $\rho_{n}^{N}(x) = \lvert \psi_{n}^{N}(x) \rvert^{2}$ probability densities and general density $\rho_{n}(x,\lambda)=\lvert\psi_{n}(x,\lambda)\rvert^{2}$ with various $\lambda$ values for $n=1\text{–}4$. We set $\mathcal{E}_{0}=x_{0}=1$ for $n=1\text{–}4$ thus $\xi=x$.}
\end{figure}

\begin{figure}
  \centering
  \includegraphics[width=\columnwidth]{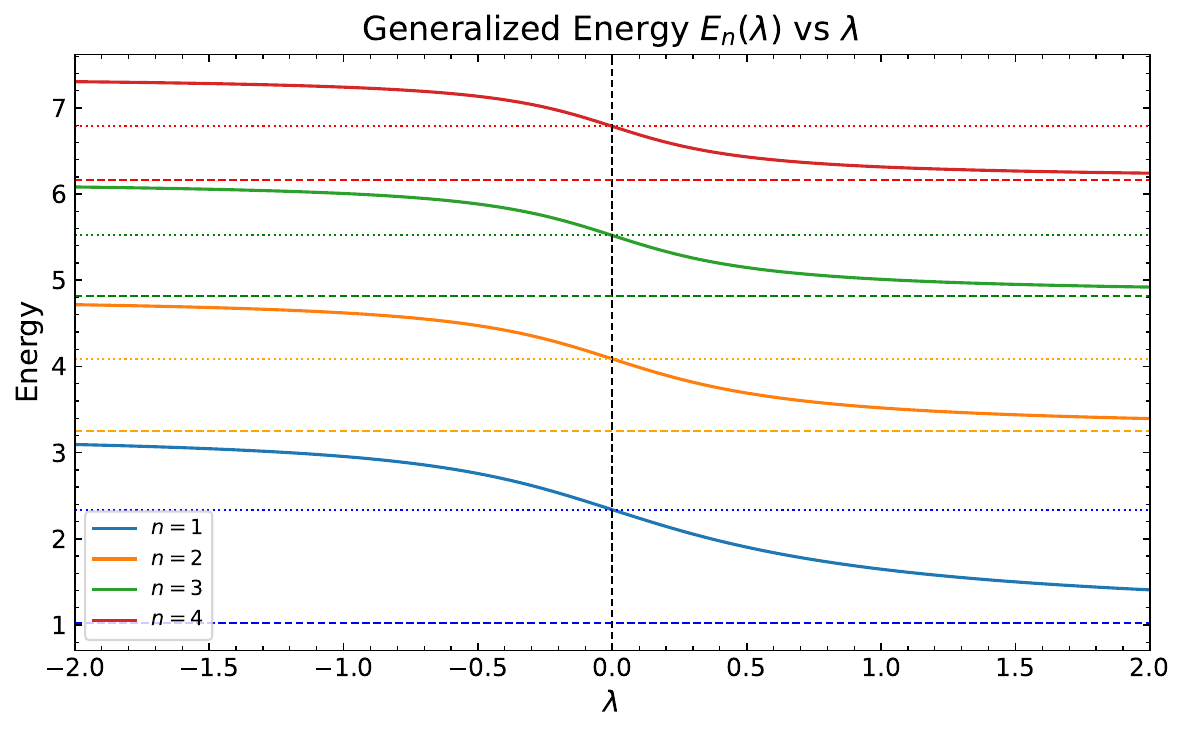}
  \caption{Generalized energies $E_{n}(
  \lambda)=-\mathcal{E}_{0} \zeta_{n}(\lambda)$ according to Eq.~\eqref{generalized_energies}. We see that the energy increases as $\lambda$ increases and vice versa. The dotted and dashed horizontal lines represent the Dirichlet $(\lambda=0)$ and Neumann $(\lambda = \infty)$ energies, respectively. We set $\mathcal{E}_{0}=x_{0}=1$ for $n=1\text{–}4$ thus $\xi=x$.}
  \label{fig:generalized_energy_plt}
\end{figure}

\subsection{Generalized Eigenfunctions}

We now use the eigenfunctions \eqref{pre_eigenfunc} of Hamiltonian \eqref{linear_pot_ham} and the general boundary condition \eqref{general_bc} to get the transcendental equation
\begin{align}
    \text{Ai}(-\varepsilon) - \lambda \text{Ai}^{\prime}(-\varepsilon) &= 0. \label{general_bc_energy_equation}
\end{align}
By numerically solving Eq.~\eqref{general_bc_energy_equation}, we find that the values of $\varepsilon$ are quantized and we denote the zeros of Eq.~\eqref{general_bc_energy_equation} as 
\begin{align}
    \zeta_{n}(\lambda), \quad n=1,2,\ldots, \label{generalized_airy_roots}
\end{align}
% where $\zeta_{n} \equiv \zeta_{n}(\lambda)$ is smoothly parameterized by the self-adjoint parameter $\lambda$ for each $n$. For consistency, we adopt the negative sign convention for $\zeta_{n}(\lambda)$, namely, $\zeta_{n}(\lambda)<0$ for each $n$. Then the quantized energies of Eq.~\eqref{general_bc_energy_equation} are
where, for consistency, we adopt the negative sign convention for $\zeta_{n}(\lambda)$, namely, $\zeta_{n}(\lambda)<0$ for each $n$. Then the quantized energies of Eq.~\eqref{general_bc_energy_equation} are
\begin{align}
    E_{n}(\lambda) &= -\mathcal{E}_{0} \zeta_{n}(\lambda), \quad n=1,2,\ldots, \label{generalized_energies}
\end{align}
and we have a complete basis of generalized eigenfunctions $\{\ket{n} \}_{n=1}^{\infty}$
\begin{align}
    \braket{x | n} &= \psi_{n}(\xi,\lambda) = \mathcal{N}_{n}(\lambda) \text{Ai} \left( \xi + \zeta_{n}(\lambda) \right), \label{generalized_eigenfunc} \\
    \mathcal{N}_{n}(\lambda) &= \abs{x_{0}\Big(\text{Ai}^{\prime}(\zeta_{n}(\lambda))^{2} - \zeta_{n}(\lambda) \text{Ai}(\zeta_{n}(\lambda))^{2} \Big)}^{-1/2}. \label{general norm const}
\end{align}
We see that the generalized energies \eqref{generalized_energies} and eigenfunctions \eqref{generalized_eigenfunc} are smoothly parameterized by the self-adjoint parameter $\lambda$ for each $n$. See Sec.~\ref{sec:exact_matrix_elements} for the derivation of the normalization constant \eqref{general norm const}.

From Fig.~\ref{fig:psigridplt} and Eq.~\eqref{general_bc_energy_equation}, we see that as $\lambda \rightarrow 0$, the generalized probability density $\rho_{n}(x,\lambda)$ converges to the Dirichlet probability density $\rho_{n}^{D}(x)$. This means that the generalized energies \eqref{generalized_energies} converge to the Dirichlet energies 
\begin{align}
    E_{n}(\lambda = 0) &=E^{D}_{n}= -\mathcal{E}_{0} a_{n}, \quad n=1,2,\ldots,
\end{align}
and the generalized eigenfunction \eqref{generalized_eigenfunc} converges to the Dirichlet eigenfunction \eqref{qbounce_eigenfunc_D}. Similarly, from Fig.~\ref{fig:psipgridplt} and Eq.~\eqref{general_bc_energy_equation}, the generalized probability density $\rho_{n}(x,\lambda)$ converges to the Neumann probability density $\rho_{n}^{N}(x)$ as $\lambda \to \infty$. This means that the generalized energies \eqref{generalized_energies} converge to the Neumann energies 
\begin{align}
    E_{n}(\lambda=\infty) &= E_{n}^{N}= -\mathcal{E}_{0} a^{\prime}_{n}, \quad n=1,2,\ldots,
\end{align}
and the generalized eigenfunction \eqref{generalized_eigenfunc} converges to the Neumann eigenfunction \eqref{qbounce_eigenfunc_N}. We see that the generalized energies \eqref{generalized_energies} shift upward for $\lambda >0$ and downward for $\lambda < 0$. It should be noted that previous work on the linear potential with the Robin condition revealed a rather rich amount of information-theoretic and thermodynamic properties \cite{olendski_theory_2016_I,olendski_theory_2016_II,olendski_thermodynamic_2018}.

\subsection{Approximate Generalized Energy Formula}\label{sec:approximate energy formulas}

In this subsection, we derive an approximation of the generalized energy \eqref{generalized_energies} using Eq.~\eqref{general_bc_energy_equation} under different $\lambda$ regimes. We consider two cases, namely, $0 <\lambda \ll 1$ and $\lambda \gg 1$. 

% The results we obtain can be viewed as the general boundary condition \eqref{general_bc} analog of the Wentzel–Kramers–Brillouin (WKB) energy formulas for the usual Dirichlet linear potential energies. 

\begin{figure}
\includegraphics[width=\columnwidth]{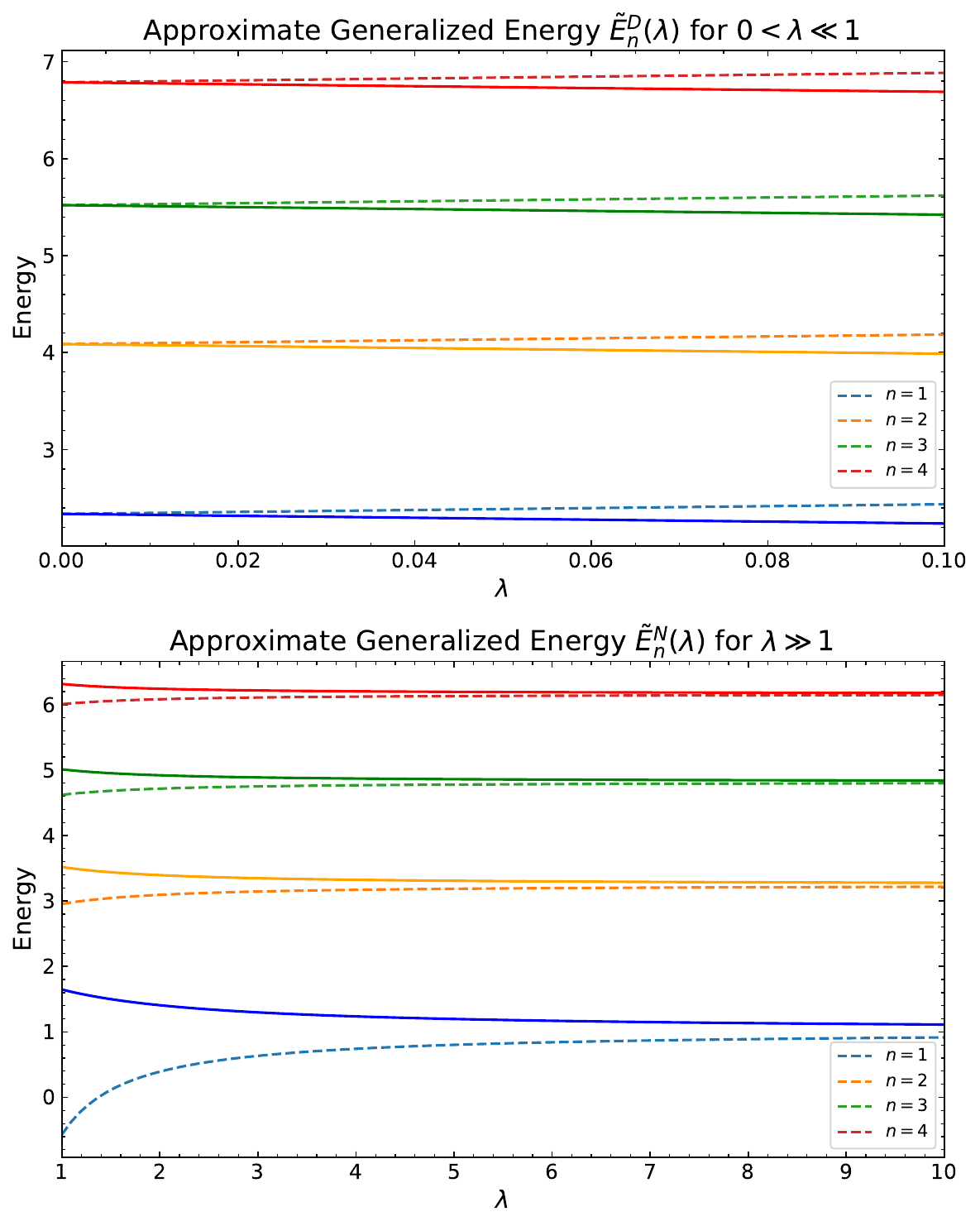}% Here is how to import EPS art
\caption{\label{fig:stacked_generalized_energy_lambda} Approximate generalized energies in the $0 <\lambda \ll 1$ and $\lambda \gg 1$ regimes corresponding to Eqs.~\eqref{dim diri case zeros} and \eqref{dim neu case zeros}, respectively. The solid and dashed lines represent the numerically calculated generalized energies \eqref{generalized_energies} and approximated energies (Eqs.~\eqref{dim diri case zeros} and \eqref{dim neu case zeros}), respectively. We set $\mathcal{E}_{0}=x_{0}=1$ for $n=1\text{–}4$ thus $\xi=x$.}
\end{figure}

\subsubsection{Dirichlet Regime \texorpdfstring{$0 < \lambda \ll 1$}{(Small Lambda)}}

From Fig.~\ref{fig:generalized_energy_plt}, we see that when $0 < \lambda \ll 1$, the generalized energies \eqref{generalized_energies} are best approximated by the Dirichlet energies \eqref{dirichlet_energy} for each $n$. Thus in the (Dirichlet) regime $0 < \lambda \ll 1$, we can expand the generalized energy $\varepsilon_{n}=\zeta_{n}$ in powers of $\lambda$ up to $O(\lambda^{4})$
\begin{align}
    \varepsilon_{n} \equiv \varepsilon_{n}(\lambda) = \varepsilon^{D}_{n} + c_{1}\lambda + c_{2}\lambda^{2} + c_{3}\lambda^{3} + O(\lambda^{4}), \label{expanded_airy_roots}
\end{align}
where $\varepsilon^{D}_{n}=a_{n}$. Now, we use Eq.~\eqref{general_bc_energy_equation} to define the function
\begin{align}
    h_{1}(x) = \text{Ai}(x) - \lambda \text{Ai}^{\prime}(x). \label{airy minus airy prime func}
\end{align}
Then we Taylor expand function \eqref{airy minus airy prime func} with Eq.~\eqref{expanded_airy_roots} to get
\begin{align}
    h_{1}(\varepsilon_{n}) &\approx \text{Ai}^{\prime}(\varepsilon^{D}_{n}) (c_{1}-1) \lambda +\text{Ai}^{\prime}(\varepsilon^{D}_{n}) c_{2} \lambda
   ^2 \notag \\
   &+\frac{1}{6} \text{Ai}^{\prime}(\varepsilon^{D}_{n}) \left(\varepsilon^{D}_{n} (c_{1}-3) c_{1}^2+6
   c_{3}\right) \lambda ^3+O\left(\lambda ^4\right),
\end{align}
where we used $\text{Ai}(\varepsilon^{D}_{n})=0$. To find the coefficients $c_{i}$ where $i=1,2,3$, recall that $h_{1}(\varepsilon_{n})=0$ thus we get
\begin{align}
    c_{1} = 1, \quad c_{2} = 0, \quad c_{3} =\frac{\varepsilon^{D}_{n}}{3}.
\end{align}
Then the approximate generalized energy in the Dirichlet regime $0 < \lambda \ll 1$ is
\begin{align}
    \varepsilon_{n}(\lambda) =\zeta_{n}(\lambda) &\approx \varepsilon^{D}_{n} \left(1 + \frac{\lambda^{3}}{3} \right) + \lambda, \label{non dim diri case zeros}
\end{align}
or, equivalently, with the correct dimensions
\begin{align}
    \tilde{E}_{n}^{D}(\lambda) &\approx E^{D}_{n} \left(1 + \frac{\lambda^{3}}{3} \right) + \mathcal{E}_{0} \lambda. \label{dim diri case zeros}
\end{align}
The energy transition formula between the $n$-th and $k$-th states is
\begin{align}
    \tilde{E}_{n,k}^{D}(\lambda) &\approx \tilde{E}_{k}^{D}(\lambda) - \tilde{E}_{n}^{D}(\lambda) =  E^{D}_{n,k} \left(1 + \frac{\lambda^{3}}{3} \right). \label{diri limit energy trans formula}
\end{align}
We see in Eq.~\eqref{diri limit energy trans formula} that, up $O(\lambda^{3})$ in the Dirichlet limit, all transition gaps are uniformly dilated. Thus, the approximate generalized spectrum \eqref{dim diri case zeros} has an $n$-independent rescaling and state-dependent shifts only enter at higher orders.

We see in Fig.~\ref{fig:stacked_generalized_energy_lambda} that Eq.~\eqref{dim diri case zeros} closely approximates the generalized energy \eqref{generalized_energies} under the Dirichlet regime. It should be noted that the famous WKB approximation of the Airy zeros \cite{wentzel_verallgemeinerung_1926, kramers_wellenmechanik_1926,Brillouin:1926blg}, namely,
\begin{align}
    \varepsilon^{D}_{n} = a_{n} \approx -\left[\frac{3\pi}{2} \left(n - \frac{1}{4} \right) \right]^{2/3},
\end{align}
can be used in Eq.~\eqref{non dim diri case zeros} in lieu of numerical computation.

\subsubsection{Neumann Regime \texorpdfstring{$\lambda \gg 1$}{(Large Lambda)}}

As can be seen in Fig.~\ref{fig:generalized_energy_plt}, we see that when $\lambda \gg 1$, the generalized energies \eqref{generalized_energies} are best approximated by the Neumann energies \eqref{neumann_energy} for each $n$. Thus in the (Neumann) regime $\lambda \gg 1$, we can expand the generalized energy $\varepsilon_{n}=\zeta_{n}$ in powers of $\kappa = 1/\lambda$ up to $O(\kappa^{4})$
\begin{align}
    \varepsilon_{n} \equiv \varepsilon_{n}(\kappa) = \varepsilon^{N}_{n} + b_{1}\kappa + b_{2}\kappa^{2} + b_{3}\kappa^{3} + O(\kappa^{4}), \label{expanded_airyp_roots}
\end{align}
where $\varepsilon^{N}_{n}=a^{\prime}_{n}$. Then we use Eq.~\eqref{general_bc_energy_equation} to define 
\begin{align}
    h_{2}(x) &= \text{Ai}^{\prime}(x) - \kappa \text{Ai}(x),
\end{align}
which we Taylor expand with Eq.~\eqref{expanded_airyp_roots} to get
\begin{align}
    h_{2}(\varepsilon_{n}) &\approx \text{Ai}(\varepsilon^{N}_{n}) (\varepsilon^{N}_{n} b_{1}+1) \kappa +\frac{1}{2}
   \text{Ai}(\varepsilon^{N}_{n}) \left(b_{1}^{2}+2 \varepsilon^{N}_{n} b_{2}\right) \kappa ^2 \notag \\
   &+\frac{1}{6}
   \text{Ai}(\varepsilon^{N}_{n}) \left((\varepsilon^{N}_{n})^{2} b_{1}^{3}+6 b_{2} b_{1}+3 \varepsilon^{N}_{n}
   \left(b_{1}^{2}+2 b_{3}\right)\right) \kappa^{3} \notag \\
   &+O\left(\kappa^4\right),
\end{align}
where we used $\text{Ai}^{\prime}(\varepsilon^{N}_{n}) = 0$. We solve for the coefficients by noting that $h_{2}(\varepsilon_{n})=0$ to get
\begin{align}
    b_{1} = \frac{1}{\varepsilon^{N}_{n}}, \quad b_{2} = -\frac{1}{2(\varepsilon^{N}_{n})^{3}}, \quad b_{3} = \frac{2(\varepsilon^{N}_{n})^{3} + 3}{6(\varepsilon^{N}_{n})^{5}}.
\end{align}
Then the approximate generalized energy in the Neumann regime $\lambda \gg 1$ ($\kappa \ll 1$) is
% \begin{align}
%     \varepsilon_{n}(\kappa) =\zeta_{n}(\kappa) &\approx \varepsilon^{N}_{n} + \frac{\kappa}{\varepsilon^{N}_{n}} -\frac{\kappa^{2}}{2(\varepsilon^{N}_{n})^{3}} + \frac{\kappa^{3}}{6}\frac{2(\varepsilon^{N}_{n})^{3} + 3}{(\varepsilon^{N}_{n})^{5}}, \label{non dim neu case zeros}
% \end{align}
\begin{align}
    \varepsilon_{n}(\lambda) &= \zeta_{n}(\lambda) \notag \\
    &\approx \varepsilon^{N}_{n} + \frac{1}{\lambda}\frac{1}{\varepsilon^{N}_{n}} - \frac{1}{\lambda^{2}}\frac{1}{2(\varepsilon^{N}_{n})^{3}} + \frac{1}{\lambda^{3}}\frac{1}{6}\frac{2(\varepsilon^{N}_{n})^{3} + 3}{(\varepsilon^{N}_{n})^{5}}, \label{non dim neu case zeros}
\end{align}
or, equivalently, with the correct dimensions
\begin{align}
    \tilde{E}_{n}^{N}(\lambda) &\approx E^{N}_{n} + \frac{1}{\lambda}\frac{\mathcal{E}^{2}_{0}}{E^{N}_{n}} - \frac{1}{\lambda^{2}}\frac{\mathcal{E}^{4}_{0}}{2(E^{N}_{n})^{3}} \notag \\
    &+ \frac{1}{\lambda^{3}}\frac{\mathcal{E}_{0}}{6} \left( \frac{2\mathcal{E}^{2}_{0}}{(E^{N}_{n})^{2}} + \frac{3\mathcal{E}^{5}_{0}}{(E^{N}_{n})^{5}} \right). \label{dim neu case zeros}
\end{align}
The energy transition formula between the $n$-th and $k$-th states is
\begin{align}
    \tilde{E}_{n,k}^{N}(\lambda) &\approx \tilde{E}_{k}^{N}(\lambda) - \tilde{E}_{n}^{N}(\lambda) \notag \\
    &=  E^{N}_{n,k} - \frac{\mathcal{E}^{2}_{0}}{\lambda} \frac{E^{N}_{n,k}}{E^{N}_{n} E^{N}_{k}} + \frac{\mathcal{E}^{4}_{0}}{2\lambda^{2}} \frac{E^{N,3}_{n,k}}{(E^{N}_{n} E^{N}_{k})^{3}} \notag \\
    &-\frac{\mathcal{E}_{0}}{6\lambda^{3}} \left( \frac{2 \mathcal{E}^{2}_{0} E^{N,2}_{n,k}}{(E^{N}_{n} E^{N}_{k})^{2}} + \frac{3 \mathcal{E}^{5}_{0} E^{N,5}_{n,k}}{(E^{N}_{n} E^{N}_{k})^{5}} \right), \label{neu limit energy trans formula}
\end{align}
where $E^{N,p}_{n,k} = (E^{N}_{k})^{p} - (E^{N}_{n})^{p}$ and $E^{N,1}_{n,k} \equiv E^{N}_{n,k}$. Unlike the Dirichlet limit \eqref{diri limit energy trans formula}, we see that the Neumann limit transition formula \eqref{neu limit energy trans formula} is state-dependent and decay as $1/\lambda$, $1/\lambda^{2}$, and $1/\lambda^{3}$. 

We see in Fig.~\ref{fig:stacked_generalized_energy_lambda} that Eq.~\eqref{dim neu case zeros} closely approximates the generalized energy \eqref{generalized_energies} under the Neumann regime. In a similar fashion, one could use the WKB approximation of the Airy prime zeros \cite{Olver:1954:AEB}
\begin{align}
    \varepsilon^{N}_{n} = a'_{n} \approx -\left[\frac{3\pi}{2} \left(n - \frac{3}{4} \right) \right]^{2/3},
\end{align}
in Eq.~\eqref{non dim neu case zeros} in lieu of numerical computation.

\section{Generalized Ehrenfest Theorems}\label{sec:generalized_ehrenfest_theorems}

In this section, we derive the Ehrenfest theorems of Hamiltonian \eqref{linear_pot_ham} with the general boundary condition \eqref{general_bc}. 

We first derive some crucial identities that will aid us in the following discussion. The general boundary condition for an arbitrary wave function $\Psi(x,t)$ is
\begin{align}
    \Psi(x,t) \Big\lvert_{x=0} &= \lambda x_{0}\frac{\partial \Psi}{\partial x} \Big\lvert_{x=0},
\end{align}
which leads to
\begin{align}
    \psi(x) \Big\lvert_{x=0} &= \lambda x_{0}\deriv{1}{x}{\psi} \Big\lvert_{x=0},
\end{align}
for the eigenfunction $\psi(x)$. From the probability current \eqref{prob current at bd}, we have that
\begin{align}
    \bc{x}{\psi^{*} \deriv{1}{x}{\psi}} = \bc{x}{\deriv{1}{x}{\psi^{*}} \psi}. \label{prime equals prime}
\end{align}
Now observe that
\begin{align}
    \bc{x}{\frac{d^{2}\psi}{dx^{2}}} &= \bc{x}{\frac{1}{x^{3}_{0}}\left(x-\varepsilon x_{0} \right)\psi} \notag \\
    &= \bc{x}{-\frac{\varepsilon}{x^{2}_{0}} \psi} \notag \\
    &= \bc{x}{\frac{\lambda x_{0}\varepsilon}{x^{2}_{0}} \deriv{1}{x}{\psi}},
\end{align}
where we used the Airy differential equation \eqref{qbounce_diff_eq} and the boundary condition \eqref{general_bc} in the first and third equalities, respectively. Then we get
% \begin{align}
%     \bc{x}{\psi^{*} \deriv{2}{x}{\psi}} &= \bc{x}{\frac{\lambda x_{0}\varepsilon}{x^{2}_{0}} \psi^{*} \deriv{1}{x}{\psi}} \notag \\
%     &= \bc{x}{\frac{\lambda x_{0}\varepsilon}{x^{2}_{0}} \deriv{1}{x}{\psi^{*}} \psi} \notag \\
%     &= \bc{x}{\deriv{2}{x}{\psi^{*}} \psi},
% \end{align}
% where we used Eq.~\eqref{prime equals prime} in the second equality. 
\begin{align}
    \bc{x}{\psi^{*} \deriv{2}{x}{\psi}} &=\bc{x}{\deriv{2}{x}{\psi^{*}} \psi}, \label{prime2 equals prime2}
\end{align}
where we used Eq.~\eqref{prime equals prime}.
% Thus
% \begin{align}
%     \bc{x}{\psi^{*} \deriv{2}{x}{\psi}} &=\bc{x}{\deriv{2}{x}{\psi^{*}} \psi}. \label{prime2 equals prime2}
% \end{align}

% {\color{red} We need to make this assumption}

Since the generalized eigenfunctions \eqref{generalized_eigenfunc} form a complete basis, we can expand an arbitrary wave function $\Psi(x,t)$ in the generalized energy basis $\ket{n}$ as
\begin{align}
    \Psi(x,t) &= \braket{x | \Psi(t)} = \sum_{n} c_{n} e^{iE_{n}t/\hbar} \psi_{n}(x), \label{expanded wavefunction}
\end{align}
where $c_{n}\equiv c_{n}(0) = \braket{n | \Psi(0)}$ are the Fourier coefficients that satisfy 
\begin{align}
    \sum_{n} \abs{c_{n}}^{2} = 1.
\end{align}
Now let $\hat{O}\equiv O(\hat{x},\hat{p})$ be an arbitrary observable. To derive the Ehrenfest theorems for $\hat{O}$ which capture the effect of the boundary \eqref{general_bc}, we take the time derivative of $\braket{\hat{O}}$ and use the Schr\"{o}dinger equation and its adjoint to get
% \begin{align}
%     \frac{d}{dt} \braket{\hat{O}} &= \frac{d}{dt} \braket{\Psi | \hat{O} | \Psi} \notag \\
%     &= \left(\frac{d}{dt} \ket{\Psi} \right)^{\dagger} \hat{O} \ket{\Psi} + \bra{\Psi} \hat{O} \left(\frac{d}{dt} \ket{\Psi} \right) \notag \\
%     &=\frac{1}{i\hbar} \left[-\braket{\Psi| \hat{H}^{\dagger} \hat{O}|\Psi} + \braket{\Psi|\hat{O} \hat{H} | \Psi} \right] \notag \\
%     &= \frac{1}{i\hbar} \left[-\braket{\hat{H}\Psi| \hat{O}\Psi} + \braket{\Psi|\hat{O} \hat{H} | \Psi}  - \braket{\Psi | \hat{H} \hat{O} | \Psi} \right. \notag \\
%     &\left. + \braket{\Psi | \hat{H} \hat{O} | \Psi} \right]\notag \\
%     &=\frac{1}{i\hbar} \left[ \braket{\Psi| [\hat{O}, \hat{H}] | \Psi} + \braket{\Psi | \hat{H} \hat{O} \Psi} - \braket{ \hat{H} \Psi |\hat{O} \Psi} \right]. 
% \end{align}
% Thus
% \begin{align}
%     \frac{d}{dt} \braket{\hat{O}} &= \frac{1}{i\hbar} \left[ \braket{\Psi| [\hat{O}, \hat{H}] | \Psi} + \braket{\Psi | \hat{H} \hat{O} \Psi} - \braket{ \hat{H} \Psi |\hat{O} \Psi} \right]. \label{general_ehrenfest_formula}
% \end{align}
\begin{align}
    \frac{d}{dt} \braket{\hat{O}} &= \frac{1}{i\hbar} \left[ \braket{\Psi| [\hat{O}, \hat{H}] | \Psi} + \braket{\Psi | \hat{H} \hat{O} \Psi} - \braket{ \hat{H} \Psi |\hat{O} \Psi} \right], \label{general_ehrenfest_formula}
\end{align}
where we have kept the surface terms generated from the integration by parts. Then we use Eq.~\eqref{expanded wavefunction} to expand the last two terms in Eq.~\eqref{general_ehrenfest_formula} to get
% In order to use the boundary condition \eqref{general_bc} in Eq.~\eqref{general_ehrenfest_formula}, we expand the last two terms in Eq.~\eqref{general_ehrenfest_formula} using Eq.~\eqref{expanded wavefunction} as follows
% \begin{align}
%     &\braket{\Psi | \hat{H} \hat{O} \Psi} - \braket{ \hat{H} \Psi |\hat{O} \Psi} \notag \\
%     &= \int^{\infty}_{0} \left[ \Psi^{\dagger}\hat{H}\hat{O} \Psi - \left(\hat{H}\Psi \right)^{\dagger} \hat{O}\Psi \right]\, dx \notag \\
%     &= \sum_{n,m} c^{\ast}_{n}(t)c_{m}(t) \int^{\infty}_{0} \left[\psi^{\dagger}_{n}\hat{H}\hat{O} \psi_{m} - \left(\hat{H}\psi_{n} \right)^{\dagger} \hat{O} \psi_{m} \right] \, dx \notag \\
%     &=\sum_{n,m} c^{\ast}_{n}(t)c_{m}(t) \left[\braket{n|\hat{H} \hat{O}m} - \braket{\hat{H}^{\dagger}n | \hat{O}m} \right] \notag \\
%     &=\sum_{n,m} c^{\ast}_{n}c_{m} e^{i\omega_{nm}t} \left[\braket{n|\hat{H} \hat{O}m} - \braket{\hat{H}^{\dagger}n | \hat{O}m} \right], \label{general ehrenfest last terms expand}
% \end{align}
\begin{align}
    &\braket{\Psi | \hat{H} \hat{O} \Psi} - \braket{ \hat{H} \Psi |\hat{O} \Psi} \notag \\
    &=\sum_{n,m} c^{\ast}_{n}c_{m} e^{i\omega_{nm}t} \left[\braket{n|\hat{H} \hat{O}m} - \braket{\hat{H}^{\dagger}n | \hat{O}m} \right], \label{general ehrenfest last terms expand}
\end{align}
where $c_{n}(t) =c_{n} e^{-i\omega_{n}t}$ and $\omega_{nm} = \omega_{n}-\omega_{m}$. With Eq.~\eqref{general ehrenfest last terms expand}, or more specifically, the inner term 
\begin{align}
    \braket{n|\hat{H} \hat{O}m} - \braket{\hat{H}^{\dagger}n | \hat{O}m}, \label{general ehrenfest last terms expand inner}
\end{align}
we are now able to use the boundary condition \eqref{general_bc}.

If we let $\hat{O}=\hat{x}$, we see that Eq.~\eqref{general ehrenfest last terms expand inner} reduces to
\begin{align}
    \braket{n|\hat{H} \hat{x}m} - \braket{\hat{H}^{\dagger}n | \hat{x}m} &=\left.\frac{\hbar^{2}}{2m} \psi^{\dagger}_n \psi_m\right\lvert_{x=0} \notag \\
    &= \frac{\hbar^{2}}{2m} \int^{\infty}_{0} \psi^{\dagger}_n \delta(x) \psi_m \, dx,
\end{align}
where we integrated by parts. Then Eq.~\eqref{general ehrenfest last terms expand} is
\begin{align}
    &\braket{\Psi | \hat{H} \hat{x} \Psi} - \braket{ \hat{H} \Psi |\hat{x} \Psi} \notag \\
    &=\frac{\hbar^{2}}{2m} \sum_{n,m} c^{\ast}_{n}(t)c_{m}(t) \int^{\infty}_{0} \psi^{\dagger}_n \delta(x) \psi_m \, dx \notag \\
    &=\frac{\hbar^{2}}{2m} \braket{\delta(\hat{x})}.
\end{align}
Similarly, if we let $\hat{O}=\hat{p}$, we get
\begin{align}
    &\braket{\Psi | \hat{H} \hat{p} \Psi} - \braket{ \hat{H} \Psi |\hat{p} \Psi} \notag \\
    &=\frac{i\hbar^{3}}{2m} \sum_{n,m}c^{\ast}_{n}c_{m} e^{i\omega_{nm}t} \left. \left[ \frac{d\psi^{\ast}_{n}}{dx} \frac{d\psi_{m}}{dx} - \psi^{\ast}_{n} \frac{d^{2}\psi_{m}}{dx^{2}} \right] \right\lvert_{x=0} \notag \\
    &= i\hbar\Braket{F_{M}\left(\hat{x}, \lambda \right)}, \label{hp term}
\end{align}
where
\begin{align}
    \Braket{F_{M}\left(\hat{x}, \lambda \right)} &= \frac{\hbar^{2}}{2m} \left. \left[ \frac{\partial\Psi^{\ast}}{\partial x} \frac{\partial\Psi}{\partial x} - \Psi^{\ast} \frac{\partial^{2}\Psi}{\partial x^{2}} \right] \right\lvert_{x=0}. \label{boundary_term_true}
\end{align}
Thus we have
\begin{align}
    \frac{d}{dt} \Braket{\hat{x}} &= \frac{1}{i\hbar}\braket{[\hat{x},\hat{H}]} \left. + \frac{\hbar}{2im}  \Psi^{\ast} \Psi \right\lvert_{x =0} \notag \\
    &= \frac{\Braket{\hat{p}}}{m} + \frac{\hbar}{2im}  \Braket{\delta(\hat{x})} \notag \\
    &\equiv \frac{\Braket{\hat{p}}}{m} -\frac{\hbar \lambda x_{0}}{2im} \Braket{\delta^{\prime}(\hat{x})}, \label{general x ehren} \\ 
    \frac{d}{dt} \Braket{\hat{p}} &= \frac{1}{i\hbar}\braket{[\hat{p},\hat{H}]} + \frac{\hbar^{2}}{2m} \left. \left[ \frac{\partial\Psi^{\ast}}{\partial x} \frac{\partial\Psi}{\partial x} - \Psi^{\ast} \frac{\partial^{2}\Psi}{\partial x^{2}} \right] \right\lvert_{x=0} \notag \\
    &= \Braket{-U^{\prime}(\hat{x})} + \Braket{F_{M}\left(\hat{x}, \lambda \right)}. \label{general p ehren}
\end{align}
We see that the general boundary condition \eqref{general_bc} has introduced nonvanishing surface terms in the Ehrenfest theorems. In the position Ehrenfest theorem \eqref{general x ehren}, a new velocity term is added. Similarly, the momentum Ehrenfest theorem \eqref{general p ehren} introduces a new force term which has been noted by previous works \cite{rokhsar_ehrenfests_1996,alonso_ehrenfest_2000,albrecht_bouncing_2023}.

Before we move on, we check if the canonical commutation relation (CCR) holds under the general boundary condition \eqref{general_bc}. Let $\varphi_{1},\varphi_{2}\in L^{2}((0,+\infty))$ be arbitrarily fixed functions that obey Eq.~\eqref{general_bc}. It can easily be shown that
\begin{align}
    \braket{[\hat{x},\hat{p}]\varphi_{2} | \varphi_{1}} &= i\hbar \braket{\varphi_{2}|\varphi_{1}}, \\
    \braket{\varphi_{2} |[\hat{x},\hat{p}] \varphi_{1}} &= i\hbar \braket{\varphi_{2}|\varphi_{1}},
\end{align}
thus
\begin{align}
    \braket{[\hat{x},\hat{p}]\varphi_{2} | \varphi_{1}} - \braket{\varphi_{2} |[\hat{x},\hat{p}] \varphi_{1}} =0.
\end{align}
Since $\varphi_{1}$ and $\varphi_{2}$ were arbitrary, we have that the CCR is self-adjoint on the half-line and has the usual value
\begin{align}
    [\hat{x},\hat{p}] = i\hbar. \label{general ccr}
\end{align}
No boundary terms appeared during the integration, thus the CCR \eqref{general ccr} is preserved even with the general boundary condition \eqref{general_bc}.

.

\section{Generalized Matrix Element Recursion} \label{sec:exact_matrix_elements}

Although the matrix elements $\braket{n|\hat{x}^{q}|k}$ and $\braket{n|\hat{p}^{q}|k}$ can be found via numerical integration with the general Airy functions \eqref{generalized_eigenfunc}, closed-form expressions can save computation time and memory. In addition, it opens up the possibility of explicitly calculating sum rules, perturbation terms, and other quantities of interest. Thus, we present the derivation of a new analytic formula of the matrix elements using the recursion methodology outlined in Refs.~\cite{goodmanson_recursion_2000,Belloni_2009}. 

We begin with the derivation of a new generalized version of the recursion relation in Ref.~\cite[Eq.~(17)]{goodmanson_recursion_2000} using Eq.~\eqref{general_bc} from which all of our results will be derived. We first simplify notation by defining
\begin{align}
    A_{n} &\equiv A_{n}(\xi,\lambda) = \text{Ai}\left(\xi + \zeta_{n}(\lambda) \right), \quad n=1,2,\ldots. \label{compact_airy_func}
\end{align}
Then the Airy differential equation can be written as
\begin{align}
    A^{\prime\prime}_{n} &= \left(\xi + \zeta_{n} \right) A_{n}, \label{compact_airy_eq}
\end{align}
where the prime represents differentiation with respect to $\xi$. Next we define a function $f(\xi)$ such that $\abs{f(0)} < \infty$ and is well behaved as $\xi \rightarrow \infty$. Then we introduce the following ansatz identity \cite{goodmanson_recursion_2000}
\begin{align}
    &\int^{\infty}_{0} d\xi \Big[ -f^{\prime\prime} (A_{n}A_{k})^{\prime} + 2f^{\prime} A^{\prime}_{n}A^{\prime}_{k} \Big]^{\prime} \notag \\
    &=\Big[-f^{\prime\prime} (A_{n}A_{k})^{\prime} + 2f^{\prime} A^{\prime}_{n}A^{\prime}_{k} \Big] \Big\lvert^{\infty}_{0}. \label{func_identity1}
\end{align}
We seek to express the left-hand side (LHS) of Eq.~\eqref{func_identity1} in terms of $A_{n}A_{k}$ since this will allow us to write the LHS in bra-ket notation. Thus, we differentiate the LHS of Eq.~\eqref{func_identity1} to get
\begin{align}
    &\int^{\infty}_{0} d\xi \Big[ -f^{\prime\prime} (A_{n}A_{k})^{\prime} + 2f^{\prime} A^{\prime}_{n}A^{\prime}_{k} \Big]^{\prime} \notag \\
    &= \int^{\infty}_{0} d\xi \Big[ -f^{\prime\prime\prime} (A_{n}A_{k})^{\prime} - 2f^{\prime\prime} A_{n}A_{k} ( 2\xi + \zeta_{n} + \zeta_{k}) \notag \\
    &+ 2f^{\prime} \left( \vphantom{\hat{\Omega}} A^{\prime}_{n}A_{k}(\xi + \zeta_{k}) + A_{n}A^{\prime}_{k} (\xi + \zeta_{n}) \right)\Big] \notag \\
    &= \int^{\infty}_{0} d\xi \, R_{1} + R_{2} + R_{3}, \label{rhs_func1}
\end{align}
where we used Eq.~\eqref{compact_airy_eq} and defined the integrands 
\begin{align}
    R_{1} &= -f^{\prime\prime\prime} (A_{n}A_{k})^{\prime}, \\
    R_{2} &= - 2f^{\prime\prime} A_{n}A_{k} ( 2\xi + \zeta_{n} + \zeta_{k}), \\
    R_{3} &= \phantom{+} 2f^{\prime} \left( \vphantom{\hat{\Omega}} A^{\prime}_{n}A_{k}(\xi + \zeta_{k}) + A_{n}A^{\prime}_{k} (\xi + \zeta_{n}) \right).
\end{align}
Integrating $R_{1}$ by parts then combining with $R_{2}$ yields
\begin{align}
    \int^{\infty}_{0} d\xi \, R_{1} + R_{2} &= -f^{\prime\prime\prime} A_{n}A_{k} \Big\lvert^{\infty}_{0} \notag \\
    &+ \int^{\infty}_{0} d\xi \, A_{n}A_{k} \Big[f^{\prime\prime\prime\prime} - 2f^{\prime\prime} (\xi + \zeta_{\text{ave}}) \Big], \label{rhs_func2}
\end{align}
where 
\begin{align}
    \zeta_{\text{ave}} = \frac{\zeta_{n} + \zeta_{k}}{2}.
\end{align}
With the identity 
\begin{align}
    ab+cd &= \frac{1}{2} (a+c)(b+d) + \frac{1}{2}(a-c)(b-d),
\end{align}
we can rewrite $R_{3}$ as
\begin{align}
    \int^{\infty}_{0} d\xi \, R_{3} &= \int^{\infty}_{0} d\xi \, \Big[ 2f^{\prime} (A_{n}A_{k})^{\prime} ( \xi + \zeta_{\text{ave}}) \notag \\
    &- f^{\prime}(A^{\prime}_{n}A_{k} - A_{n}A^{\prime}_{k}) (\zeta_{n} - \zeta_{k}) \Big]. \label{rhs_func3}
\end{align}
We integrate each term in Eq.~\eqref{rhs_func3} by parts with Eq.~\eqref{compact_airy_eq} to get 
\begin{align}
    \int^{\infty}_{0} d\xi \, R_{3} &= \Big[2f^{\prime} (\xi + \zeta_{\text{ave}}) A_{n}A_{k} \notag \\
    &- f(\zeta_{n} - \zeta_{k})(A^{\prime}_{n}A_{k} - A_{n}A^{\prime}_{k}) \Big]\Big\lvert^{\infty}_{0} \notag \\
    &+ \int^{\infty}_{0} d\xi \, A_{n}A_{k} \Big[ -2f^{\prime\prime} (\xi+ \zeta_{\text{ave}}) - 2f^{\prime} \notag \\
    &+ f(\zeta_{n} - \zeta_{k})^{2} \Big],
\end{align}
which we combine with Eq.~\eqref{rhs_func2} to get for the LHS of Eq.~\eqref{func_identity1}
\begin{align}
    &\int^{\infty}_{0} d\xi \, R_{1} + R_{2} + R_{3} \notag \\
    &= \Big[ \left( \vphantom{\hat{\Omega}} -f^{\prime\prime\prime} + 2f^{\prime} (\xi + \zeta_{\text{ave}}) \right) A_{n}A_{k} \notag \\
    &- f(\zeta_{n} - \zeta_{k})(A^{\prime}_{n}A_{k} - A_{n}A^{\prime}_{k}) \Big] \Big\lvert^{\infty}_{0} \notag \\
    &+ \int^{\infty}_{0} d\xi \, A_{n}A_{k} \Big[ f^{\prime\prime\prime\prime} - 4f^{\prime\prime} (\xi + \zeta_{\text{ave}})  - 2f^{\prime} \notag \\
    &+ f(\zeta_{n} - \zeta_{k})^{2} \Big]. \label{general_func_identity_LHS}
\end{align}
Finally, we move the boundary terms in Eq.~\eqref{general_func_identity_LHS} to the right-hand side (RHS) of Eq.~\eqref{func_identity1} to get our desired recursion relation identity
\begin{align}
    &\int^{\infty}_{0} d\xi \, A_{n}A_{k} \Big[ f^{\prime\prime\prime\prime} - 4f^{\prime\prime} (\xi + \zeta_{\text{ave}}) - 2f^{\prime} + f(\zeta_{n} - \zeta_{k})^{2} \Big] \notag \\
    &=\Big[-f^{\prime\prime} (A_{n}A_{k})^{\prime} + 2f^{\prime} A^{\prime}_{n}A^{\prime}_{k} \notag \\
    &+ \left( \vphantom{\hat{\Omega}} f^{\prime\prime\prime} - 2f^{\prime} (\xi + \zeta_{\text{ave}}) \right) A_{n}A_{k} \notag \\
    &+ f(\zeta_{n} - \zeta_{k})(A^{\prime}_{n}A_{k} - A_{n}A^{\prime}_{k})\Big] \Big\lvert^{\infty}_{0}. \label{general_func_identity}
\end{align}

Before expressing Eq.~\eqref{general_func_identity} in bra-ket notation, we first consider the special case $n=k$ and $f(\xi) = \xi$ which yields
\begin{align}
    \int^{\infty}_{0} d\xi \, A^{2}_{n} &= \Big[ (\xi + \zeta_{n}) A^{2}_{n} - (A^{\prime}_{n})^{2} \Big] \Big\lvert^{\infty}_{0} \notag \\
    &=\left[ A^{\prime}_{n}(0, \lambda)^{2} - \zeta_{n}(\lambda) A_{n}(0, \lambda)^{2} \right]. \label{pre_norm1}
\end{align}
We see that Eq.~\eqref{pre_norm1} is the normalization constant $\mathcal{N}_{n}(\lambda)$ (sans $x_{0}$) thus we have
\begin{align}
    \mathcal{N}_{n}(\lambda) &= \abs{x_{0} \Big(A^{\prime}_{n}(0, \lambda)^{2} - \zeta_{n}(\lambda) A_{n}(0, \lambda)^{2} \Big)}^{-1/2} \notag \\
    &\equiv \abs{x_{0}\Big(\text{Ai}^{\prime}(\zeta_{n}(\lambda))^{2} - \zeta_{n}(\lambda) \text{Ai}(\zeta_{n}(\lambda))^{2} \Big)}^{-1/2}, \label{norm1}
\end{align}
where $\mathcal{N}_{n}(\lambda)$ depends on both $n$ and $\lambda$. We see that $\mathcal{N}_{n}(\lambda)$ reduces to the Dirichlet \eqref{diri norm} and Neumann \eqref{neu norm} normalization constants in the limits $\lambda \to 0$ and $\lambda \to \infty$, respectively. Then the generalized eigenfunctions \eqref{generalized_eigenfunc} in our simplified notation \eqref{compact_airy_eq} are
\begin{align}
    \Braket{x | n} &= \psi_{n}(\xi,\lambda) = \mathcal{N}_{n}(\lambda) A_{n}(\xi,\lambda). \label{general_eigenfunc_with_An}
\end{align}
To further aid our next discussion, we note that due to the oscillatory behavior of the Airy function for negative arguments (recall that $\zeta_{n}$ is negative), the sign of $\text{Ai}(\zeta_{n})$ and $\text{Ai}^{\prime}(\zeta_{n})$ alternates with $n$. Thus, if we define the following non-negative quantities 
\begin{align}
    \alpha^{(0)}_{n} &\equiv \alpha_{n}(\lambda) = \mathcal{N}_{n}(\lambda) \abs{A_{n}(0, \lambda)}, \notag \\
    \alpha^{(1)}_{n} &\equiv \alpha^{\prime}_{n}(\lambda) = \mathcal{N}_{n}(\lambda) \abs{A^{\prime}_{n}(0, \lambda)}, \notag \\
    & \vdotswithin{ \equiv } \notag \\
    \alpha^{(k)}_{n} &\equiv \alpha^{(k)}_{n}(\lambda) = \mathcal{N}_{n}(\lambda) \abs{A^{(k)}_{n}(0, \lambda)}, \label{alpha_alphap_defs}
\end{align}
where $k = 0,1,\ldots$ denotes the order of differentiation, we can compactly write $\psi_{n}(0,\lambda)$ and $\psi_{n}'(0,\lambda)$ as
\begin{align}
    \psi_{n}(0,\lambda) &= (-1)^{n+1} \alpha_{n}(\lambda), \label{psi_0_ident} \\
    \psi^{\prime}_{n}(0,\lambda) &= (-1)^{n+1} \alpha^{\prime}_{n}(\lambda), \label{psi_1_ident}
\end{align}
respectively. In addition, it can easily be shown using Eqs.~\eqref{general_bc}, \eqref{norm1}, and \eqref{general_eigenfunc_with_An} that
\begin{align}
    \alpha_{n}(0) &= 0, \label{alpha_zero} \\
    \alpha^{\prime}_{n}(0) &= (-1)^{n+1}, \label{alpha_prime_zero} \\
        \left(\alpha^{\prime}_{n} \right)^{2} - \zeta_{n}  \left(\alpha_{n} \right)^{2} &= 1, \label{alpha_sum}
\end{align}
for $n=1,2,\ldots$. In particular, Eq.~\eqref{alpha_sum} will prove useful in evaluating the matrix elements as well as the sum rules in the next section.

To write Eq.~\eqref{general_func_identity} in bra-ket notation, we let $f(\xi) = \xi^{q}$, where $q$ is a nonnegative integer, and replace $A_{n}$ with the generalized eigenfunction \eqref{general_eigenfunc_with_An} on the LHS. On the RHS of Eq.~\eqref{general_func_identity}, we replace $A_{n}(0,\lambda)$ and $A_{n}'(0,\lambda)$ with Eqs.~\eqref{psi_0_ident} and \eqref{psi_1_ident}, respectively, after evaluating the limit at $+\infty$. Then Eq.~\eqref{general_func_identity} becomes
\begin{widetext}
\begin{align}
    &q(q-1)(q-2)(q-3) \braket{n| \hat{\xi}^{q-4} |k} - 4q(q-1)\zeta_{\text{ave}}\braket{n| \hat{\xi}^{q-2} |k} - 2q(2q-1)\braket{n| \hat{\xi}^{q-1} |k} + (\zeta_{n} - \zeta_{k})^{2} \braket{n | \hat{\xi}^{q} |k} \notag \\
    &= (-1)^{n+k} \left[q(q-1) (\alpha^{\prime}_{n}\alpha_{k} + \alpha_{n} \alpha^{\prime}_{k}) \delta_{2q} - 2q\alpha^{\prime}_{n}\alpha^{\prime}_{k} \delta_{1q} - \left( q(q-1)(q-2)\delta_{3q} - 2q\zeta_{\text{ave}}\delta_{1q} \right)\alpha_{n}\alpha_{k} \right], \label{braket_general_func_identity}
\end{align}
\end{widetext}
where we used Eq.~\eqref{psi_0_ident}-\eqref{psi_1_ident}, $\delta_{ab}$ is the Kronecker delta function, and it is understood that matrix elements with negative powers of $\hat{\xi}$ are ignored since the LHS of Eq.~\eqref{braket_general_func_identity} was obtained via differentiation of $\hat{\xi}^{q}$. When $n=k$, the recursion formula \eqref{braket_general_func_identity} simplifies to
\begin{widetext}
\begin{align}
    &q(q-1)(q-2)(q-3) \braket{n| \hat{\xi}^{q-4} |n} - 4q(q-1)\zeta_{n}\braket{n| \hat{\xi}^{q-2} |n} - 2q(2q-1)\braket{n| \hat{\xi}^{q-1} |n}  \notag \\
    &= 2q(q-1) \alpha^{\prime}_{n}\alpha_{n} \delta_{2q} - 2q(\alpha^{\prime}_{n})^{2} \delta_{1q} - \left( q(q-1)(q-2)\delta_{3q} - 2q\zeta_{n}\delta_{1q} \right)\alpha_{n}^{2}. \label{braket_n_equals_n_func_identity}
\end{align}
\end{widetext}

With Eqs.~\eqref{braket_general_func_identity} and \eqref{braket_n_equals_n_func_identity}, we can now derive the matrix elements of $\hat{\xi}^{q}$ which are
\begin{widetext}

\begin{empheq}[left={n = k:\empheqlbrace}]{alignat=3}
    &q=1: \quad \braket{n|n} &&= \phantom{-} 1, \label{x_elem1_diag} \\
    &q=2: \quad \braket{n| \hat{\xi} |n} &&= -\frac{1}{3} ( \alpha_{n}\alpha^{\prime}_{n} + 2\zeta_{n}), \label{x_elem2_diag} \\
    &q=3: \quad \braket{n| \hat{\xi}^{2} |n} &&= \phantom{-} \frac{1}{5} \left[ \alpha^{2}_{n} + \frac{4}{3}\zeta_{n} \left(\alpha_{n}\alpha^{\prime}_{n} + 2\zeta_{n} \right) \right], \phantom{(-1)^{n+k+1}\left[\frac{24}{(\zeta_{n} - \zeta_{k})^{4}} (\alpha^{\prime}_{n}\alpha^{\prime}_{k} - \zeta_{\text{ave}} \alpha_{n}\alpha_{k})  \right]} \label{x_elem3_diag} 
    % &q=4: \quad \braket{n| \hat{\xi}^{3} |n} &&= \frac{3}{7} \left[1 - \frac{2}{5} \zeta_{n} \left( \alpha^{2}_{n} + \frac{4}{3}\zeta_{n} \left(\alpha_{n}\alpha^{\prime}_{n} + 2\zeta_{n} \right) \right) \right], \label{x_elem4_diag} \\
    % &q=5: \quad \braket{n| \hat{\xi}^{4} |n} &&=  \label{x_elem5_diag} 
\end{empheq}
\begin{empheq}[left={n \neq k:\empheqlbrace}]{alignat=3}
    &q=0: \quad \braket{n|k} &&= \phantom{-} 0, \label{x_elem0_nondiag} \\
    &q=1: \quad \braket{n| \hat{\xi} |k} &&= \phantom{-} \frac{2(-1)^{n+k+1}}{(\zeta_{n} - \zeta_{k})^{2}}(\alpha^{\prime}_{n}\alpha^{\prime}_{k} - \zeta_{\text{ave}} \alpha_{n}\alpha_{k}), \label{x_elem1_nondiag} \\
    &q=2: \quad \braket{n| \hat{\xi}^{2} |k} &&= \phantom{-} (-1)^{n+k+1}\left[\frac{24}{(\zeta_{n} - \zeta_{k})^{4}} (\alpha^{\prime}_{n}\alpha^{\prime}_{k} - \zeta_{\text{ave}} \alpha_{n}\alpha_{k}) - \frac{2}{(\zeta_{n} - \zeta_{k})^{2}} (\alpha^{\prime}_{n}\alpha_{k} + \alpha_{n}\alpha^{\prime}_{k}) \right], \label{x_elem2_nondiag} 
    % &q=3: \quad \braket{n| \hat{\xi}^{3} |k} &&= \label{x_elem5_nondiag} \\
    % &q=4: \quad \braket{n| \hat{\xi}^{4} |k} &&= \label{x_elem6_nondiag} 
\end{empheq}

\end{widetext}
We see from Eqs.~\eqref{x_elem1_diag} and \eqref{x_elem0_nondiag} that Eq.~\eqref{braket_general_func_identity} verifies the orthonormality of our generalized eigenkets $\ket{n}$. Note that since $\hat{\xi}=\hat{x}/x_{0}$, we can restore the units by multiplying $x^{q}_{0}$ to Eqs.~\eqref{x_elem1_diag}-\eqref{x_elem2_nondiag}. Further closed-form formulas of $\braket{n|\hat{x}^{q}|k}$ for higher $q$ can be found using Eqs.~\eqref{braket_general_func_identity} and \eqref{braket_n_equals_n_func_identity}.

To find the matrix elements of $\hat{p}^{2}$, we use the identity
\begin{align}
    \hat{p}^{2} &= 2m (\hat{H} - F_{0}\hat{x}), \label{p2_identity}
\end{align}
with $E_{n}=-\mathcal{E}_{0}\zeta_{n}$ then sandwich Eq.~\eqref{p2_identity} with the generalized energy eigenkets $\ket{n}$ and $\ket{k}$. To find the matrix elements of $\hat{p}$, one could integrate by parts $\braket{n|\hat{p}|k}$ and use Eqs.~\eqref{qbounce_diff_eq} and \eqref{general_bc}, but we opt for a simpler method using the Ehrenfest theorem. Recall that the typical Ehrenfest theorem for $\hat{x}$ is
\begin{align}
    \Braket{\hat{p}} &= m \frac{d}{dt} \Braket{\hat{x}}, \label{ehrenfest_x_normal}
\end{align}
but the boundary condition \eqref{general_bc} will introduce nonzero surface terms in the RHS of Eq.~\eqref{ehrenfest_x_normal}, hence spoiling the self-adjointness of $\hat{p}$. Thus we use the general Ehrenfest theorem formula \eqref{general_ehrenfest_formula} with $\hat{O} = \hat{x}$ to get
% \begin{align}
%     \Braket{\hat{p}} &= m\frac{d}{dt} \Braket{\hat{x}} \notag \\
%     &= \frac{m}{i\hbar} \left. \braket{[\hat{x}, \hat{H} ]} + \frac{\hbar}{2i} \left[ x\left( \frac{\partial\Psi^{\ast}}{\partial x} \Psi - \Psi^{\ast} \frac{\partial\Psi}{\partial x} \right) - \Psi^{\ast} \Psi  \right] \right\lvert^{\infty}_{0} \notag \\
%     &= \left. \frac{m}{i\hbar}  \braket{[\hat{x}, \hat{H}]} + \frac{\hbar}{2i}  \Psi^{\ast} \Psi  \right\lvert_{x=0} \notag \\
%     &= \frac{m}{i\hbar}  \braket{[\hat{x}, \hat{H} ]} + \frac{\hbar}{2i} \braket{\delta(\hat{x})}. \label{momentum ehrenfest matrix elem}
% \end{align}
\begin{align}
    \Braket{\hat{p}} &= m\frac{d}{dt} \Braket{\hat{x}} = \frac{m}{i\hbar}  \braket{[\hat{x}, \hat{H} ]} + \frac{\hbar}{2i} \braket{\delta(\hat{x})}. \label{momentum ehrenfest matrix elem}
\end{align}
Since Eq.~\eqref{momentum ehrenfest matrix elem} holds for any arbitrary state $\ket{\Psi(t)}$, we can remove the averaging to get the following operator identity
\begin{align}
     \hat{p} &= \frac{m}{i\hbar}  [\hat{x}, \hat{H} ] + \frac{\hbar}{2i} \delta(\hat{x}) = \frac{mx_{0}}{i\hbar}  [\hat{\xi}, \hat{H} ] + \frac{\hbar}{2ix_{0}} \delta(\hat{\xi}). \label{p_op_identity}
\end{align}
Sandwiching Eq.~\eqref{p_op_identity} with the generalized energy eigenkets $\ket{n}$ and $\ket{k}$ then yields
\begin{align}
    \Braket{n | \hat{p} | k} &= -\frac{i\hbar}{2x_{0}} \left[\left(\zeta_{n} - \zeta_{k} \right) \braket{n| \hat{\xi} |k} + \braket{n | \delta(\hat{\xi}) | k} \right]. \label{general p matrix elem formula}
\end{align}
Gathering our results leads to
\begin{widetext}

\begin{empheq}[left={n = k:\empheqlbrace}]{alignat=3}
&\braket{n| \hat{p} |n} &&= - \frac{i\hbar}{2x_{0}} \alpha^{2}_{n}, \label{p_elem1_diag} \\
&\braket{n| \hat{p}^{2} |n} &&= -\frac{2}{3} m\mathcal{E}_{0} \left(\zeta_{n} - \alpha_{n}\alpha^{\prime}_{n} \right), \phantom{\frac{1}{(\zeta_{n} - \zeta_{k})} \left(\alpha^{\prime}_{n}\alpha^{\prime}_{k} - \zeta_{\text{ave}} \alpha\alpha_{k}\right) + \frac{\alpha_{n} \alpha_{k}}{2}} \label{p_elem2_diag}
\end{empheq}
\begin{empheq}[left={n \neq k:\empheqlbrace}]{alignat=3}
&\braket{n| \hat{p} |k} &&= \frac{i\hbar}{x_{0}} (-1)^{n+k+1}\left[- \frac{1}{(\zeta_{n} - \zeta_{k})} \left(\alpha^{\prime}_{n}\alpha^{\prime}_{k} - \zeta_{\text{ave}} \alpha_{n}\alpha_{k}\right) + \frac{1}{2} \alpha_{n} \alpha_{k} \right], \label{p_elem1_nondiag} \\
&\braket{n| \hat{p}^{2} |k} &&= -\frac{4 m\mathcal{E}_{0} (-1)^{n+k+1}}{(\zeta_{n} - \zeta_{k})^{2}}(\alpha^{\prime}_{n}\alpha^{\prime}_{k} - \zeta_{\text{ave}} \alpha_{n}\alpha_{k}). \label{p_elem2_nondiag}
\end{empheq}

% \begin{empheq}[left={\eqmath[r]{A}{n=k:}\empheqlbrace}]{alignat=2}
% &\eqmath[l]{B}{\begin{aligned}
% &\braket{n| \hat{p} |n} = 0, 
% \end{aligned}} \\
% &\eqmath[l]{B}{\begin{aligned}
% &\braket{n| \hat{p}^{2} |n} = -\frac{2}{3} m\mathcal{E}_{0}( \zeta_{n} - \alpha_{n}\alpha^{\prime}_{n}),
% \end{aligned}} 
% \end{empheq}
% \begin{empheq}[left={\eqmath[r]{A}{n\neq k:}\empheqlbrace}]{alignat=2}
% &\eqmath[l]{B}{\begin{aligned}
% &\braket{n| \hat{p} |k} = -\frac{i\hbar}{x_{0}} \frac{(-1)^{n+k+1}}{(\zeta_{n} - \zeta_{k})}(\alpha^{\prime}_{n}\alpha^{\prime}_{k} - \zeta_{\text{ave}} \alpha_{n}\alpha_{k}),
% \end{aligned}} \\
% &\eqmath[l]{B}{\begin{aligned}
% &\braket{n| \hat{p}^{2} |k} = -\frac{4 m\mathcal{E}_{0} (-1)^{n+k+1}}{(\zeta_{n} - \zeta_{k})^{2}}(\alpha^{\prime}_{n}\alpha^{\prime}_{k} - \zeta_{\text{ave}} \alpha_{n}\alpha_{k}).
% \end{aligned}} 
% \end{empheq}

\end{widetext}
We note that the matrix elements of $\hat{p}$ are complex, but this is to be expected since $\hat{p}$ is not self-adjoint under the boundary condition \eqref{general_bc}, which means that $\hat{p}$ will not admit a real spectrum. Compared to the Dirichlet case, our diagonal matrix elements \eqref{p_elem1_diag}-\eqref{p_elem2_diag} of $\hat{p}$ possess nonzero entries for $\lambda \neq 0$. 

Numerical evaluation on \textsc{Mathematica} and Python demonstrate excellent agreement with our exact matrix element formulas. It can easily be shown by using Eqs.~\eqref{alpha_zero}-\eqref{alpha_sum} that when $\lambda = 0$, the results of this section, i.e. Eqs.~\eqref{braket_general_func_identity}-\eqref{x_elem2_nondiag} and Eqs.~\eqref{general p matrix elem formula}-\eqref{p_elem2_nondiag}, reduce to their Dirichlet equivalents in Refs.~\cite{goodmanson_recursion_2000, Belloni_2009}. It should be noted that the general boundary condition roots $\zeta_{n}$ are the only quantity that needs numerical approximation for our matrix element formulas. In lieu of numerical computation, the approximate generalized energy formulas \eqref{non dim diri case zeros} and \eqref{non dim neu case zeros} from Sec.~\ref{sec:approximate energy formulas} serve as an excellent substitute.

\section{Generalized Sum Rules}\label{sec:generalized_sum_rules}

In many areas of physics, sum rules reveal key insights, such as forbidden transitions and exact expressions for perturbation terms. In Sec.~\ref{sec:exact_matrix_elements}, we saw that the matrix elements of $\hat{x}$ and $\hat{p}$ were changed significantly due to the general boundary condition \eqref{general_bc}. It immediately follows that the sum rules for the linear potential are expected to change. In this section, we derive the generalized Thomas–Reiche–Kuhn (TRK) sum rule as well as other sum rules.

There are many sum rules that exist across many fields, thus we choose to focus on some of the most commonly used and recognized sum rules. Observe that for any observable $O(\hat{x})$ and $k\geq1$, we have \cite{jackiw_quantum-mechanical_1967,wang_generalization_1999}
\begin{align}
    &\sum_{m} (E_{m} - E_{n})^{k} \abs{\braket{n|O(\hat{x})|m}}^{2} \notag \\
    &= \braket{n|O(\hat{x})\underbrace{[\hat{H}, [\hat{H},\ldots,[\hat{H}}_{k \text{ times}},O(\hat{x})],\ldots]]|n}, 
\end{align}
which can be simplified to
\begin{align}
    &\sum_{m} (E_{m} - E_{n})^{k} \abs{\braket{n|O(\hat{x})|m}}^{2} \notag \\
    &= \left(-\frac{i\hbar}{2m}\right)^{\!k}
    \sum_{j=0}^k\binom{k}{j} \braket{n|O(\hat{x}) \hat{p}^{j} O^{(k)}(\hat{x}) \hat{p}^{k-j} |n} \notag \\
    &+(k-1) \left(-\frac{i\hbar}{2m}\right)^{k-2} \frac{\hbar^{2} F_{0}}{m} \braket{n|O(\hat{x}) O^{(k-1)}(\hat{x})|n}.
\end{align}
When $k=0$, we have
\begin{align}
    \sum_{m} \abs{\braket{n|O(\hat{x})|m}}^{2} = \braket{n|O(\hat{x})^{2}|n}.
\end{align}
For $O(\hat{x})=\hat{x}^{q}$, we have
\begin{align}
    &\sum_{m} (E_{m} - E_{n})^{k} \abs{\braket{n|\hat{x}^{q}|m}}^{2} \notag \\
    &= \Bigl(-\frac{i\hbar}{2m}\Bigr)^{k} (q)_k \sum_{j=0}^{k}\binom{k}{j} \braket{n|\hat{x}^{q} \hat{p}^{j} \hat{x}^{q-k} \hat{p}^{k-j}|n} \notag \\
    &+ (k-2) \Bigl(-\frac{i\hbar}{2m}\Bigr)^{k-2}\frac{\hbar^{2}F_{0}}{m}(q)_{k-1} \braket{n|\hat{x}^{2q-k+1}|n},
\end{align}
where $(q)_{n}=q(q-1)\cdots (q-n+1)$ is the falling factorial. Then we have
\begin{widetext}
    \begin{alignat}{2}
        &k=0, q=1: \quad &&\sum_{m} \abs{\braket{n|\hat{x}|m}}^{2} = \braket{n|\hat{x}^{2}|n} =\frac{x_{0}^{2}}{5} \left[ \alpha^{2}_{n} + \frac{4}{3}\zeta_{n} \left(\alpha_{n}\alpha^{\prime}_{n} + 2\zeta_{n} \right) \right], \label{k0 q1} \\
        &k=1, q=1: \quad &&\sum_{m} (E_{m} - E_{n}) \abs{\braket{n|\hat{x}|m}}^{2} = \frac{1}{2}\braket{n|[\hat{x},[\hat{H},\hat{x}]]|n} = \frac{\hbar^{2}}{2m}, \label{k1 q1} \\
        &k=1, q=2: \quad &&\sum_{m} (E_{m} - E_{n}) \abs{\braket{n|\hat{x}^{2}|m}}^{2} = \frac{1}{2}\braket{n|[\hat{x}^{2},[\hat{H},\hat{x}^{2}]]|n} = \frac{2\hbar^{2}}{m} \braket{n|\hat{x}^{2}|n} = \frac{2\hbar^{2}x_{0}^{2}}{5m} \left[ \alpha^{2}_{n} + \frac{4}{3}\zeta_{n} \left(\alpha_{n}\alpha^{\prime}_{n} + 2\zeta_{n} \right) \right], \label{k1 q2} \\
        &k=2, q=1: \quad &&\sum_{m} (E_{m} - E_{n})^{2} \abs{\braket{n|\hat{x}|m}}^{2} = \braket{n|[\hat{H},\hat{x}]^{\dagger} [\hat{H},\hat{x}]|n} = \frac{\hbar^{2}}{m^{2}} \braket{n|\hat{p}^{2}|n} = -\frac{2\hbar^{2}}{3m} \mathcal{E}_{0} \left(\zeta_{n} - \alpha_{n}\alpha^{\prime}_{n} \right). \label{k2 q1} 
    \end{alignat}
\end{widetext}
Eqs.~\eqref{k0 q1}-\eqref{k1 q2} are the closure/completeness, TRK (dipole), and monopole sum rules, respectively. 
% Inserting the matrix elements then gives
% \begin{widetext}
%     \begin{alignat}{2}
%         &k=0, q=1: \quad &&\sum_{m} \abs{\braket{n|\hat{x}|m}}^{2} = \braket{n|\hat{x}^{2}|n} =\frac{x_{0}^{2}}{5} \left[ \alpha^{2}_{n} + \frac{4}{3}\zeta_{n} \left(\alpha_{n}\alpha^{\prime}_{n} + 2\zeta_{n} \right) \right], \label{k0 q1 s sum} \\
%         &k=1, q=1: \quad &&\sum_{m \neq n} (E_{m} - E_{n}) \abs{\braket{n|\hat{x}|m}}^{2} = \braket{n|[\hat{H},\hat{x}]|n} = -\frac{i\hbar}{m} \braket{n|\hat{p}|n} = - \frac{\hbar^{2}}{2mx_{0}} \alpha^{2}_{n}, \label{k1 q1 s sum} \\
%         &k=1, q=2: \quad &&\sum_{m \neq n} (E_{m} - E_{n}) \abs{\braket{n|\hat{x}^{2}|m}}^{2} = \frac{1}{2}\braket{n|[\hat{x}^{2},[\hat{H},\hat{x}^{2}]]|n} = \frac{2\hbar^{2}}{m} \braket{n|\hat{x}^{2}|n}, \label{k1 q2 s sum} \\
%         &k=2, q=1: \quad &&\sum_{m \neq n} \frac{1}{(\zeta_{n} - \zeta_{k})^{2}}(\alpha^{\prime}_{n}\alpha^{\prime}_{k} - \zeta_{\text{ave}} \alpha_{n}\alpha_{k})^{2} = \frac{\hbar^{2}}{4mF_{0}x_{0}^{2}}. \label{k2 q1 s sum} 
%     \end{alignat}
% \end{widetext}
We can also calculate the Bethe sum rule
\begin{align}
    \sum_{m} (E_{m} - E_{n}) \abs{\braket{n|e^{iq\hat{x}}|m}}^{2} &= \frac{1}{2}\braket{n|[e^{-iq\hat{x}},[\hat{H},e^{iq\hat{x}}]]|n} \notag \\
    &= \frac{\hbar^{2} q^{2}}{2m},
\end{align}
which remains unchanged.

As we observed in the last section, our generalized sum rules reduce to their Dirichlet versions in the limit $\lambda \to 0$ as seen in Refs.~\cite{goodmanson_recursion_2000, Belloni_2009}. Similarly, the Neumann versions can be recovered in the limit $\lambda \to \infty$. Thus, our results up to now are the most general results of the linear potential \eqref{linear_pot_ham}.

\section{Generalized Heisenberg Uncertainty relation}\label{sec:generalized_heisenberg_uncertainty_rel}

In this section, we use the results of Sec.~\ref{sec:exact_matrix_elements} to find the Heisenberg uncertainty relation with the general boundary condition \eqref{general_bc}. 

% Before we calculate the uncertainty relation, we first check if the CCR holds under the general boundary condition. Let $\varphi_{1},\varphi_{2}\in L^{2}((0,\infty))$ be arbitrarily fixed functions that obey Eq.~\eqref{general_bc}. It can easily be shown that
% \begin{align}
%     \braket{[\hat{x},\hat{p}]\varphi_{2} | \varphi_{1}} &= i\hbar \braket{\varphi_{2}|\varphi_{1}}, \\
%     \braket{\varphi_{2} |[\hat{x},\hat{p}] \varphi_{1}} &= i\hbar \braket{\varphi_{2}|\varphi_{1}},
% \end{align}
% thus
% \begin{align}
%     \braket{[\hat{x},\hat{p}]\varphi_{2} | \varphi_{1}} - \braket{\varphi_{2} |[\hat{x},\hat{p}] \varphi_{1}} =0.
% \end{align}
% Since $\varphi_{1}$ and $\varphi_{2}$ were arbitrary, we have that the CCR is self-adjoint and has the usual value
% \begin{align}
%     [\hat{x},\hat{p}] = i\hbar.
% \end{align}
% Thus, the CCR is preserved even with the general boundary condition \eqref{general_bc}. 

Let $\hat{A}$ and $\hat{B}$ be operators that are not necessarily self-adjoint, with the variance defined as
\begin{align}
    (\Delta\hat{A})^{2} = \braket{\hat{A}^{\dagger} \hat{A}} - \braket{\hat{A}^{\dagger}} \braket{\hat{A}}.
\end{align}
Then the generalized uncertainty relation for $\hat{A}$ and $\hat{B}$ is \cite{albrecht_bouncing_2023}
\begin{align}
    \Delta\hat{A} \Delta\hat{B} \geq \abs{\braket{\hat{A}^{\dagger} \hat{B}} - \braket{\hat{A}^{\dagger}} \braket{\hat{B}}}. \label{general uncertainty}
\end{align}
If we let $\hat{A}=\hat{x}$ and $\hat{B}=\hat{p}$, we have
\begin{align}
    (\Delta\hat{x})_{n} (\Delta\hat{p})_{n} \geq \abs{\braket{\hat{x} \hat{p}}_{n} - \braket{\hat{x}}_{n} \braket{\hat{p}}_{n}}. \label{general uncert rel 0}
\end{align}
where $\braket{\cdot}_{n} \equiv \braket{n|\cdot|n}$ denotes averaging with respect to the energy basis \eqref{generalized_eigenfunc}. To evaluate Eq.~\eqref{general uncert rel 0}, observe that
\begin{align}
    \braket{\{\hat{x},\hat{p}\}}_{n} &= \sum_{k} \left[\braket{n|\hat{x}|k} \braket{k|\hat{p}|n} + \braket{n|\hat{p}|k} \braket{k|\hat{x}|n} \right] \notag \\
    &= \frac{2i\hbar}{x_{0}} \sum_{k} \frac{\alpha_{n}' \alpha_{k}'}{(\zeta_{n} - \zeta_{k})^{2}}(\alpha_{n}'\alpha_{k}' - \zeta_{\text{ave}} \alpha_{n}\alpha_{k}). \label{anti comm xp}
\end{align}
To solve Eq.~\eqref{anti comm xp}, note that
\begin{align}
    \frac{d}{dt} \braket{\hat{x}^{2}} = \frac{1}{i\hbar} \braket{[\hat{x}^{2}, \hat{H}]} = \frac{1}{m} \braket{\{\hat{x},\hat{p}\}}, 
\end{align}
where we used Eq.~\eqref{general_ehrenfest_formula} and integrated by parts using the general boundary condition \eqref{general_bc}. Since the time derivative vanishes in a stationary state, the sum in Eq.~\eqref{anti comm xp} must equal zero and we get
% Since $\{\hat{x},\hat{p}\}$ is self-adjoint, the sum in Eq.~\eqref{anti comm xp} must equal zero due to the fact that self-adjointness guarantees a real spectrum, thus 
\begin{align}
    \braket{\{\hat{x},\hat{p}\}}_{n} = 0.
\end{align}
Then we use the CCR \eqref{general ccr} to get
\begin{align}
    \braket{\hat{x} \hat{p}}_{n} &= \frac{1}{2} \braket{\{\hat{x},\hat{p} \}}_{n} + \frac{i\hbar}{2} = \frac{i\hbar}{2}.
\end{align}
With Eqs.~\eqref{x_elem2_diag} and \eqref{p_elem1_diag}, Eq.~\eqref{general uncert rel 0} evaluates to
\begin{align}
    (\Delta\hat{x})_{n} (\Delta\hat{p})_{n} \geq \frac{\hbar}{6} \abs{3-\alpha_{n}^{2}(\lambda)\left(\alpha_{n}(\lambda) \alpha_{n}'(\lambda) + 2\zeta_{n}(\lambda) \right)}. \label{general uncertainty eigenfunc form}
\end{align}
We see that the generalized Heisenberg uncertainty relation \eqref{general uncertainty eigenfunc form} yields a new lower bound that is smoothly parameterized by $\lambda$. In addition, we see that Eq.~\eqref{general uncertainty eigenfunc form} reduces to the usual $\hbar/2$ lower bound for both the Dirichlet ($\lambda \to 0$) and Neumann ($\lambda \to \infty$) cases. It should be noted that previous works \cite{al-hashimi_particle_2012,al-hashimi_self-adjoint_2012,albrecht_bouncing_2023} derived similar generalized uncertainty relations as well.

Although it looks as if the general uncertainty relation \eqref{general uncertainty eigenfunc form} violates the Heisenberg uncertainty relation, this is not the case. Recall that the Heisenberg uncertainty relation applies to only observables, but the momentum operator $\hat{p}$ on the half-line is not self-adjoint therefore it is not an observable in this context. Instead, the general uncertainty relation \eqref{general uncertainty eigenfunc form} serves as a fingerprint of the boundary effects, rather than a minimal uncertainty limit for observables.

% In previous works \cite{al-hashimi_particle_2012,al-hashimi_self-adjoint_2012,albrecht_bouncing_2023}, similar generalized uncertainty relations were derived for a wide range of problems. 
Note that Heisenberg mused (see page 30 of Ref.~\cite{heisenberg2013physical}) about the fact that the uncertainty relation needs to change its form if additional constraints are specified.

% \begin{subfigure}[t]{.49\textwidth}
%         \includegraphics[width=\columnwidth]{generalized_diri_prob_density_plot.pdf}% Here is how to import EPS art
%         \caption{\label{fig:psigridplt} General probability density $\rho_{n}(x,\lambda)$ in the limit $\lambda \rightarrow 0$. Note how $\rho_{n}(x,\lambda)$ starts to fit $\rho_{n}^{D}(x)$ with the Dirichlet condition as $\lambda \rightarrow 0$.}
%     \end{subfigure}

\begin{figure}
\includegraphics[width=\columnwidth]{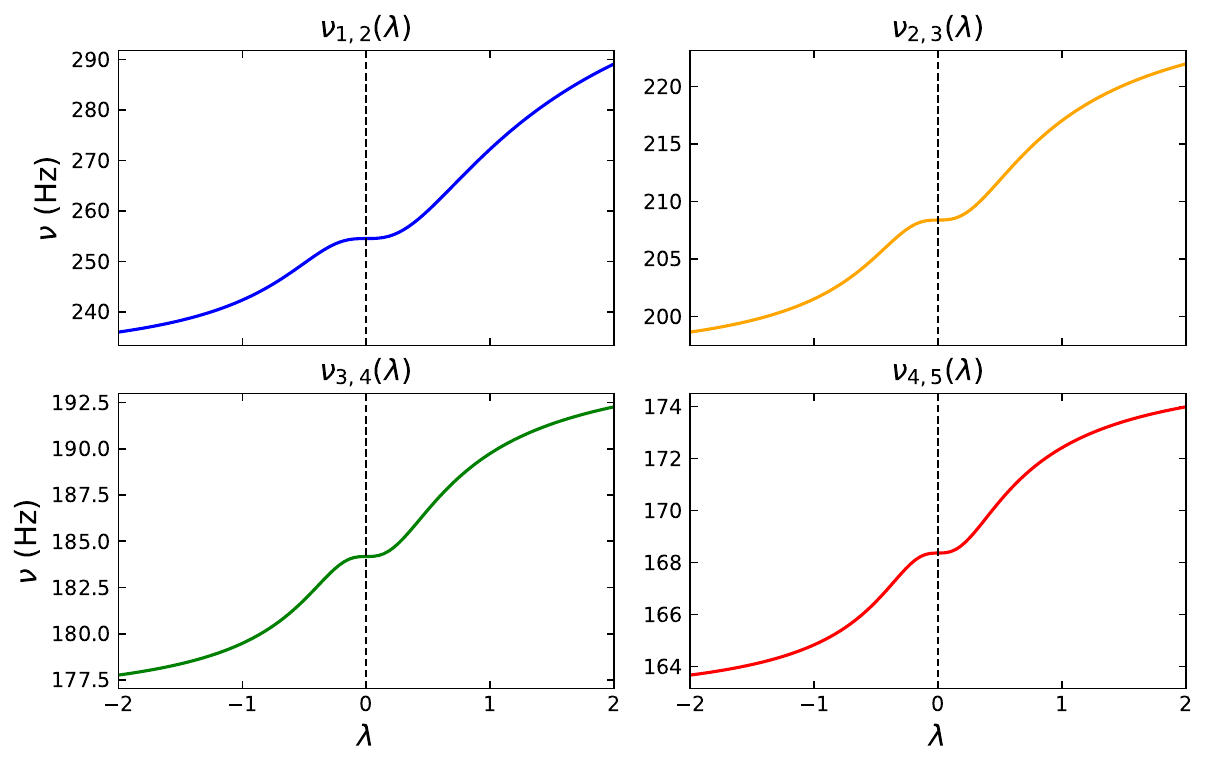}% Here is how to import EPS art
\caption{\label{fig:fourvdiffplot} Transition frequency $\nu_{n,n+1}(\lambda)=\nu_{n+1}(\lambda) - \nu_{n}(\lambda)$ as a function of the self-adjoint parameter $\lambda$. The vertical dashed lines are the transition frequencies with the Dirichlet condition. We use $g_{c}=9.804925 \text{ m}/\text{s}^{2}$ thus $\mathcal{E}_{0} = 0.6016 \, \text{peV}$.}
\end{figure}

\begin{figure}
\centering
\includegraphics[width=\columnwidth]{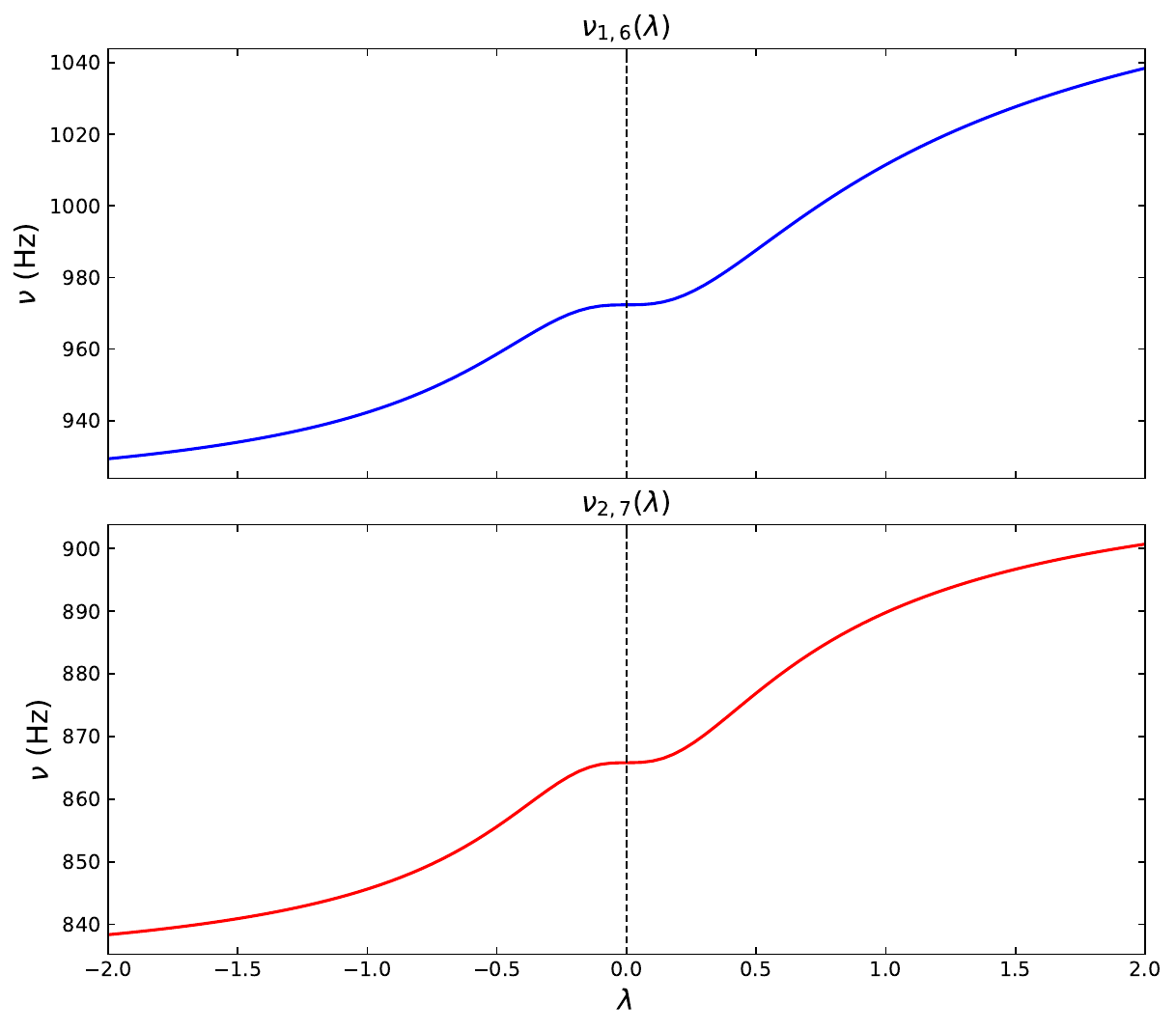}% Here is how to import EPS art
\caption{\label{fig:vdiffplt50mod} Transition frequencies $\nu_{1,6}(\lambda)$ and $\nu_{2,7}(\lambda)$ as a function of the self-adjoint parameter $\lambda$. The vertical dashed lines are the transition frequencies with the Dirichlet condition. We use $g_{c}=9.804925 \text{ m}/\text{s}^{2}$ thus $\mathcal{E}_{0} = 0.6016 \, \text{peV}$.}
\end{figure}

\section{Application to UltraCold Neutron Gravitational Interferometry} \label{sec:applications}

% \subsection{UltraCold Neutron Gravitational Interferometry}

A typical application of the linear potential \eqref{linear_pot_ham} is for free-falling UCN experiments \cite{nesvizhevsky_quantum_2002,nesvizhevsky_measurement_2003,cronenberg_acoustic_2018,micko_qbounce_2023,mickoQBounceRamseySpectroscopy2023}. In this set up, $F_{0}=mg_{c}$ and the potential is
\begin{align}
    U(x) = \begin{cases}
        mg_{c}x, &\quad x\geq 0, \\
        \infty, &\quad x<0,
    \end{cases}
\end{align}
where $m$ is the (neutron) mass and $g_{c}$ is the local gravitational acceleration at the ILL. This gives a characteristic length and energy value of $x_{0} = 5.87 \, \mu \text{m}$ and $\mathcal{E}_{0} = 0.6016 \, \text{peV}$, respectively. In many of these experiments, UCNs in a gravitational field free-fall onto a vibrating mirror, which drives the quantum state transitions of the neutrons. The mirror is often modeled as an infinite barrier that keeps neutrons from penetrating into the mirror. This leads to utilizing the Dirichlet boundary condition for modeling the short-range interaction of the neutrons with the mirror. This approximation, albeit a very accurate one, suffices for such experiments due to the low-energy nature of UCNs and their low penetrability, but realistically, some penetration of UCNs into the mirror is to be expected. This penetration is already observed and accounted for in experiments \cite{chizhova_vectorial_2014}. The most general boundary condition to account for this is Eq.~\eqref{general_bc}.

\begin{figure}
\includegraphics[width=\columnwidth]{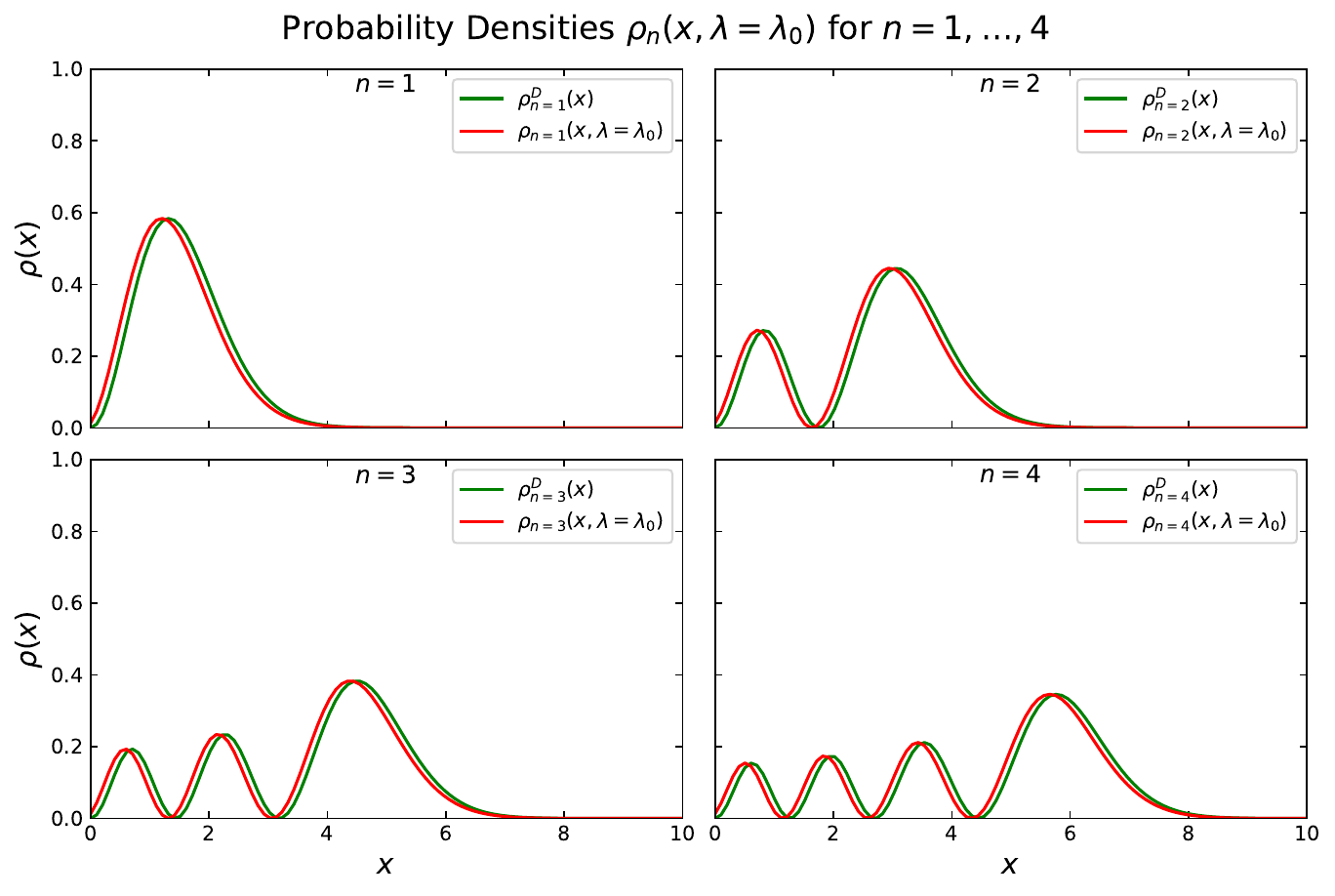}% Here is how to import EPS art
\caption{\label{fig:seladjeigflampltgrid} First few energy levels of the generalized probability density \eqref{self_adj_prob_density_lamb0} $\rho_{n}(x,\lambda_{0})$ compared to the Dirichlet probability density $\rho^{D}_{n}(x)$. We see that $\rho_{n}(x,\lambda_{0})$ is slightly translated to the left compared to $\rho^{D}_{n}(x)$. We set $\mathcal{E}_{0}=x_{0}=1$ for $n=1\text{–}4$ thus $\xi=x$.}
\end{figure}

There have been some experiments where experimental data did not match expected theoretical predictions. For instance, the authors of \cite{altarawneh_spatial_2024} found that the experimental data of \cite{ichikawa_observation_2014} did not definitively correspond to the spatial distribution of quantum gravitational states of UCNs. In a more recent \qb experiment \cite{micko_qbounce_2023,mickoQBounceRamseySpectroscopy2023} performed at the instrument
PF2 at ILL, the $\ket{1} \to \ket{6}$ transition frequency of UCNs was measured and yielded 
\begin{align}
    \overline{\nu}_{1,6} &= 972.842 \pm 0.0456057 \, \text{Hz} \notag \\
    &= 972.842(045) \, \text{Hz}, \label{qb_trans_freq_16} 
\end{align}
where the bar $\overline{\nu}_{k,n} =\nu_{n} - \nu_{k}$ denotes the experimental value. The measured transition frequency \eqref{qb_trans_freq_16} corresponded to a local gravitational acceleration value of \cite[Eq.~(6.1)]{mickoQBounceRamseySpectroscopy2023}
\begin{align}
    g &= g_{\rm qB} \pm \sigma g_{\rm qB} - \delta g_{\rm sys} \pm \sigma g_{\rm sys} \notag \\
    &= 9.81253 \pm 6.9 \cdot 10^{-4} - 6.1 \cdot 10^{-4} \pm 6.5 \cdot 10^{-4} \, \text{m}/\text{s}^{2} \notag \\
    &= 9.81192(95) \, \text{m}/\text{s}^{2}. \label{qb_g_16} 
\end{align}
For our work, the self-adjoint parameter $\lambda$ may play a role in the observed energy shift \eqref{qb_trans_freq_16}. To find the value of $\lambda = \lambda_{0}$ that reproduces the frequency shift \eqref{qb_trans_freq_16}, and its local gravitational acceleration value \eqref{qb_g_16}, we can minimize the following chi-squared function
\begin{align}
    \chi^2(\lambda_{0}) &= \left(\frac{
        2\pi\hbar\overline{\nu}_{1,6} + \mathcal{E}_{0} [\zeta_{6}(\lambda_{0}) - \zeta_{1}(\lambda_{0})]
    }{2\pi\hbar\Delta\overline{\nu}_{1,6}}\right)^2, \label{chi squared fitting func}
\end{align}
to get
\begin{align}
    \lambda_{0} &= \lambda_{\text{min}} \pm \Delta\lambda_{0} = 0.11928 \pm 0.00995, \label{lambda0 value}
    % \lambda_{0} &= \lambda_{\text{min}} \pm \Delta\lambda_{0} = 0.11732 \pm 0.00260, \label{lambda0 value}
    % \lambda_{0} &= \lambda_{\text{min}} \pm \Delta\lambda_{0} = 0.11654 \pm 0.00027, \label{lambda0 value}
    % \lambda_{0} &= \lambda_{\text{min}} \pm \Delta\lambda_{0} = 0.11799 \pm 0.00026, \label{lambda0 value}
    % \lambda_{0} &= \lambda_{\text{min}} \pm \Delta\lambda_{0} = 0.11799(26), \label{lambda0 value}
    % \lambda_{0} &= \lambda_{\text{min}} \pm \Delta\lambda_{0} = 0.117988991 \pm 0.00025788, \label{lambda0 value}
    % \lambda_{0} &= \lambda_{\text{min}} \pm \Delta\lambda_{0} = 0.11803 \pm 0.00025, \label{lambda0 value}
    % \lambda_{0} &= \lambda_{\text{min}} \pm \Delta\lambda_{0} = 0.11658 \pm 00026
\end{align}
where $\Delta\overline{\nu}_{1,6}$ and $\Delta\lambda_{0}$ are the errors. We note that the $\ket{2} \to \ket{7}$ transition was measured as well but the $\ket{1} \to \ket{6}$ was the statistically dominant measurement \cite{micko_qbounce_2023,mickoQBounceRamseySpectroscopy2023}. Thus our chi-squared fitting \eqref{chi squared fitting func} suffices as a first impression. This gives
\begin{align}
    \braket{x | n,\lambda_{0}} &= \psi_{n}(\xi,\lambda_{0}) = \mathcal{N}_{n}(\lambda_{0}) \text{Ai} \left( \xi + \zeta_{n}(\lambda_{0}) \right), \label{self_adj_eig_lamb0} \\
    \rho_{n}(x,\lambda_{0}) &= \abs{\psi_{n}(\xi,\lambda_{0})}^{2}, \label{self_adj_prob_density_lamb0} \\
    E_{n}(\lambda_{0}) &= -\mathcal{E}_{0}\zeta_{n}(\lambda_{0}), \label{self_adj_energy_lamb0}
\end{align}
with Tab.~\ref{tab:freqgrid} showing the first few energy levels \eqref{self_adj_energy_lamb0} compared to the usual Dirichlet energies \eqref{dirichlet_energy} along with their difference. In Fig.~\ref{fig:seladjeigflampltgrid}, we show the generalized probability density \eqref{self_adj_prob_density_lamb0} at $\lambda_{0}$ compared to the Dirichlet probability density $\rho_{n}^{D}(x)$. Since $\lambda_{0}$ is positive, this physically means that the mirror is more repulsive than the usual Dirichlet mirror. Thus, we observe an upward shift in the transition frequencies (see Fig.~\ref{fig:vdiffplt50mod}).

With our estimated value of $\lambda_{0}$, we can calculate the probability $P_{\rm in}^{(n)}(\lambda_{0})$ of finding a neutron in the $n$-th state under the mirror. Since the generalized eigenfunction \eqref{generalized_eigenfunc} is not defined for $x\in (-\infty,0)$, we elect to use the exponentially decaying wave
\begin{align}
    \Phi_{n}(x,\lambda_{0}) = e^{\kappa_{0} x} \psi_{n}(0,\lambda_{0}),
\end{align}
as an approximate wave function for under the mirror. To find the decay constant $\kappa_{0}$, we use the general boundary condition \eqref{general_bc_units} to get the state-independent
\begin{align}
    \kappa_{0} = \frac{\psi'(0)}{\psi(0)} = \frac{1}{\lambda_{0}x_{0}} = 1428451.34430 \text{ m}^{-1}.
\end{align}
Then the decaying wave function is
\begin{align}
    \Phi_{n}(x,\lambda_{0}) &= e^{\kappa_{0} x} \psi_{n}(0,\lambda_{0}) = e^{x/(\lambda_{0} x_{0})} \psi_{n}(0,\lambda_{0}),
\end{align}
and the probability under the mirror for state $n$ is
% \begin{align}
%     l_{0} &= \lambda_{0} x_{0} = 688.562 \pm 15.287 \, \rm{nm} = 689 (15) \, \rm{nm}.
% \end{align}
% \begin{align}
%     l_{0} &= \lambda_{0} x_{0} = 688.562 \pm 15.287 \, \rm{nm} = 689 (15) \, \rm{nm}.
% \end{align}
\begin{align}
    P_{\rm in}^{(n)}(\lambda_{0}) &= \int^{0}_{-\infty} \abs{\Phi_{n}(x,\lambda_{0})}^{2} \, dx \notag \\
    &= \int^{0}_{-\infty} e^{2x/(\lambda_{0}x_{0})} \abs{\psi_{n}(0,\lambda_{0})}^{2} \, dx \notag \\
    &= x_{0}\int^{0}_{-\infty} e^{2\xi/\lambda_{0}} \abs{\psi_{n}(0,\lambda_{0})}^{2} \, d\xi \notag \\
    &= \frac{\lambda_{0}x_{0}}{2}\abs{\psi_{n}(0,\lambda_{0})}^{2}. \label{penet prob calc}
\end{align}
For the $n=1$ state, we have
\begin{align}
    P_{\rm in}^{(1)} &=0.00082. \label{penet prob}
\end{align}
Our calculation \eqref{penet prob} represents an estimate.

Our calculation of $\lambda_{0}$ is a rough approximation, as the latest \qb experiment measured neutron transmissions and subsequently calculated transition frequencies and local gravitational acceleration values. In addition, our model \eqref{linear_pot_ham} did not account for the systematic effects, such as the Earth's rotation, that affected the measurement of the neutron's transition frequency in the \qb experiment \cite{micko_qbounce_2023,mickoQBounceRamseySpectroscopy2023}. This means that the self-adjoint parameter $\lambda$ is, in general, an \textit{effective} parameter that ``soaks'' up these unaccounted contributions in addition to quantifying boundary effects. By adding these systematic effects into our model \eqref{linear_pot_ham}, we can refine $\lambda_{0}$ even further since a more accurate model would lead to a smaller $\lambda_{0}$ which approximates the Dirichlet boundary condition. Nevertheless, our value of $\lambda_{0}$ remains reasonable because the observed small energy shift corresponds to a small $\lambda_{0}$ value. 
%We therefore expect Eq.~\eqref{lambda0 value} to lie close to the true value of $\lambda_{0}$. 
To refine Eq.~\eqref{lambda0 value} further, we plan to perform a numerical simulation of the \qb experiment with the reported systematic effects to extract $\lambda_{0}$ more precisely, which we will detail in a follow-up paper.

Since the value of the self-adjoint parameter $\lambda$ also depends on the composition of the mirror material, one can use $\lambda$ as a free tunable parameter. %This opens the door to reanalyze the interpretation of discrepancies in neutron transmission. In addition, it gives the possibility of ruling out exotic effects, such as chameleon fields \cite{khouryChameleonFieldsAwaiting2004,braxStronglyCoupledChameleons2011,ivanovInfluenceChameleonField2013,cronenberg_acoustic_2018} and dark energy \cite{hinterbichlerScreeningLongRangeForces2010,hinterbichlerSymmetronCosmology2011,jenkeGravityResonanceSpectroscopy2014,altarawneh_ultra-cold_2025}, that are theorized to occur on the boundary of the mirror, which are predicted to give small contributions to neutron counts. Since the wave function of the UCNs is highly sensitive to the precision of the generalized Airy roots \eqref{generalized_airy_roots} and the $\lambda$ value, a small $\lambda$ could potentially affect the interpretation of any unknown short-range forces that operate within the characteristic length scales. By repeating free-falling UCN experiments using mirrors with different coatings or compositions, the value of $\lambda$ can be determined and used to remove noise from measurements systematically. 

\begin{table}
    \centering
    \begin{tabular}{@{\extracolsep{4pt}} || c | c | c | c||}
        \hline 
        $n$ & $E^{D}_{n}$ \text{(peV)} & $E_{n} (\lambda_0)$ \text{(peV)} & $\delta E_{n}$ \text{(peV)} \\
        \hline
        1 & 1.4066 & 1.3356 & 0.0710 \\
        \hline
        2 & 2.4592 & 2.3888 & 0.0704 \\
        \hline
        3 & 3.3211 & 3.2511 & 0.0700 \\
        \hline
        4 & 4.0827 & 4.0131 & 0.0696 \\
        \hline
        5 & 4.7790 & 4.7098 & 0.0692 \\
        \hline
        6 & 5.4278 & 5.3590 & 0.0689 \\
        \hline
        7 & 6.0400 & 5.9713 & 0.0686 \\
        \hline
    \end{tabular}
    \caption{First few energy levels (in peV) of the Dirichlet $E^{D}_{n}$ \eqref{dirichlet_energy} and generalized $E_{n}(\lambda_{0})$ \eqref{self_adj_energy_lamb0} energies along with their difference $\delta E_{n} = E^{D}_{n} - E_{n}(\lambda_{0})$. We use $g_{c}=9.804925 \text{ m}/\text{s}^{2}$ $\mathcal{E}_{0} = 0.6016 \, \text{peV}$ for each of the energies.}
    \label{tab:freqgrid}
\end{table}

% \textcolor{blue}{With $\lambda_{0}$, we can now predict the $\nu_{1,3}$ transition frequency and its associated local gravitational acceleration value.}

% \begin{table}[]
%     \centering
%     \begin{tabular}{@{\extracolsep{4pt}} || c | c | c | c ||}
%         \hline 
%         $n \rightarrow n+1$ & $\Delta\nu_{n+1,n}$ \text{(Hz)} & $\Delta\nu_{n+1,n} (\lambda_0)$ \text{(Hz)} & $\delta \nu_{n+1,n}$ \text{(Hz)}\\
%         \hline %        $1\rightarrow2$ & 254.56345 & 254.69143 & -0.12798 \\
%         \hline
%         $2\rightarrow3$ & 208.41324 & 208.51545 & -0.10221 \\
%         \hline
%         $3\rightarrow4$ & 184.19667 & 184.28516 & -0.08849 \\
%         \hline
%         $4\rightarrow5$ & 168.37990 & 168.45933 & -0.07943 \\
%         \hline
%         $5\rightarrow6$ & 156.90049 & 156.97329 & -0.07280 \\
%         \hline
%         $6\rightarrow7$ & 148.02724 & 148.09488 & -0.0676 \\
%         \hline 
%     \end{tabular}
%     \caption{Transition frequencies of \eqref{dirichlet_energy} and \eqref{self_adj_energy_lamb0} where $\delta \nu_{n+1,n} = \Delta\nu_{n+1,n} - \Delta\nu_{n+1,n} (\lambda _0)$. Note that the transition frequency $\Delta\nu_{n+1,n}(\lambda_{0})$ shifts downward and the strength of $\lambda_{0}$ decreases for higher $n$.}
%     \label{tab:freqgridVdiff}
% \end{table}

\section{Discussion and Outlook} \label{sec:discussion and outlook}

We applied the theory of self-adjoint extensions to the linear potential Hamiltonian \eqref{linear_pot_ham} on the half-line and obtained the most general boundary condition \eqref{general_bc} under which $\hat{H}$ is self-adjoint. We generalized existing results of the Dirichlet linear potential to the general boundary condition version by deriving new formulas for the energy, eigenfunction, matrix elements, and various sum rules. We demonstrated that local gravitational acceleration measurements can be influenced by these nontrivial boundary effects. We provided a method to increase the sensitivity of the \qb experiment in addition to proposing that hypothetical fifth forces could be potentially ruled out as nothing more than boundary effects. %Also, we provided other potential applications in other areas of UCN physics. 
Our work highlights that the notion of self-adjointness and boundary conditions must be taken into strict consideration when modeling physical systems of interest. 

As mentioned in the previous section, we plan to provide a follow up paper that numerically simulates the recent \qb experiment \cite{micko_qbounce_2023,mickoQBounceRamseySpectroscopy2023} in order to extract the precise value of the self-adjoint parameter $\lambda$. From the results of this work and previous works, we see that the self-adjoint parameter might affect not just the different $g$ measurements of \cite{micko_qbounce_2023,mickoQBounceRamseySpectroscopy2023}, but also open the door to new applications. % such as those mentioned in Sec.~\ref{sec:applications}
More importantly, since the self-adjoint parameter $\lambda$ is a tunable parameter, our results open the possibility of quantum control via boundary engineering \cite{balmasedaQuantumControlBoundary2019}. Given the wide range of fields in which self-adjoint extensions apply and the recent advances in experimental design and control across those fields, it might be only a matter of time before self-adjoint extensions are experimentally verified.

% Despite the application to free-falling UCNs, our results of the generalized linear potential can be applied to many other fields, such as those mentioned in the introduction, by choosing an appropriate value of $F_{0}$ in Eq.~\eqref{linear_pot_ham}. More importantly, since the self-adjoint parameter $\lambda$ is a tunable parameter, our results open the possibility of quantum control via boundary engineering. Some examples include the linear Stark effect and semiconductor heterojunctions \cite{al-hashimi_particle_2012}.

% For condensed matter systems, the self-adjoint parameter $\lambda$ depends on the specific material and can be dependent on position ($\lambda \equiv \lambda(x)$) \cite{al-hashimi_particle_2012}.

\begin{acknowledgments}

E.J.S. was supported by the ARO Undergraduate Research Apprenticeship Program. D.I.B. was also supported by Army Research Office (ARO) (grant W911NF-23-1-0288; program manager Dr.~James Joseph). H.A. has received funding from the Austrian Research Promotion Agency (FFG), Project Number 896034 and 921409. The views and conclusions contained in this document are those of the authors and should not be interpreted as representing the official policies, either expressed or implied, of ARO, or the U.S. Government. The U.S. Government is authorized to reproduce and distribute reprints for Government purposes notwithstanding any copyright notation herein. 

\end{acknowledgments}

\appendix
\renewcommand{\theequation}{\thesection\arabic{equation}}

\section{Proof of \texorpdfstring{$\lambda \in \mathbb{R}$}{TEXT}} \label{app:proof_of_lambda_in_real}

In this section, we prove that
\begin{align}
    \frac{\text{Ai}^{\prime}(i\varepsilon_{\eta}) + \alpha^{\ast} \text{Ai}^{\prime}(-i\varepsilon_{\eta})}{\text{Ai}(i\varepsilon_{\eta}) + \alpha^{\ast} \text{Ai}(-i\varepsilon_{\eta})} &= \lambda \in \mathbb{R},
\end{align}
where $\alpha \in U(1)$ and $\varepsilon_{\eta} > 0$. 

Fix an arbitrary $\alpha \in U(1)$. Since $\text{Ai}(x)$ and $\text{Ai}^{\prime}(x)$ functions are analytic, we have that
\begin{align}
    \left(\text{Ai}(ix) \right)^{\ast} &= \text{Ai}(-ix), \quad \left(\text{Ai}^{\prime}(ix) \right)^{\ast} = \text{Ai}^{\prime}(-ix),
\end{align}
for any $x \in \mathbb{R}$. Then we have
\begin{align}
    \lambda^{\ast} &= \frac{\text{Ai}^{\prime}(-i\varepsilon_{\eta}) + \alpha \text{Ai}^{\prime}(i\varepsilon_{\eta})}{\text{Ai}(-i\varepsilon_{\eta}) + \alpha \text{Ai}(i\varepsilon_{\eta})} \left( \frac{\alpha^{\ast}}{\alpha^{\ast}} \right) \notag \\
    &= \frac{\text{Ai}^{\prime}(i\varepsilon_{\eta}) + \alpha^{\ast} \text{Ai}^{\prime}(-i\varepsilon_{\eta})}{\text{Ai}(i\varepsilon_{\eta}) + \alpha^{\ast} \text{Ai}(-i\varepsilon_{\eta})} = \lambda.
\end{align}
Thus, $\lambda^{*}=\lambda$ and since $\alpha$ and $x$ were arbitrary, $\lambda$ is a real constant.

\bibliography{boundary_condition}% Produces the bibliography via BibTeX.

\end{document}

%% file: preamble.tex
\usepackage{amsthm}
\usepackage{mathtools,amssymb,bm,bbm}
\usepackage{tensor}
\usepackage{microtype}
\usepackage{braket}
\usepackage{dcolumn}
\usepackage{xcolor}
\usepackage{graphicx}
\usepackage[left=23mm,right=13mm,top=35mm,columnsep=15pt]{geometry} 
\usepackage{adjustbox}
\usepackage{placeins}
\usepackage[T1]{fontenc}
\usepackage{lipsum}
\usepackage{csquotes}
\usepackage{slashed}
\usepackage{xspace}